\documentclass[pdflatex,sn-mathphys-num]{sn-jnl}

\usepackage{graphicx}%
\usepackage{multirow}%
\usepackage{amsmath,amssymb,amsfonts}%
\usepackage{amsthm}%
\usepackage{mathrsfs}%
\usepackage[title]{appendix}%
\usepackage{xcolor}%
\usepackage{textcomp}%
\usepackage{manyfoot}%
\usepackage{booktabs}%
\usepackage{algorithm}%
\usepackage{algorithmicx}%
\usepackage{algpseudocode}%
\usepackage{listings}%

\theoremstyle{thmstyleone}%
\theoremstyle{thmstyletwo}%

\theoremstyle{thmstylethree}%

\begin{document}

\title[Article Title]{The basic ideas of quantum mechanics}


\author{\fnm{David} \sur{Ellerman}}

\affil{\orgdiv{https://orcid.org/0000-0002-5718-618X}, \orgaddress{Independent researcher}, \city{Ljubljana}, \state{Slovenia}, \country{david@ellerman.org}}


\abstract{For a century, quantum theorists have been reading the mathematical entrails of quantum mechanics (QM) to divine the nature of quantum reality. But to little avail. In this paper a different approach is taken, namely to identify and explain the basic intuitive ideas involved in QM. This does not tell one how those basic `gears' all mesh together in the beautiful mathematics of QM. But this does give one some intuitive (\textit{anschaulich})) ideas about the quantum reality described in the seemingly hard-to-interpret mathematical framework.}




\maketitle
\tableofcontents
\section{Introduction: A new approach}\label{sec1}

For a century, quantum theorists have been reading the mathematical entrails
of quantum mechanics (QM) to divine the nature of quantum reality. But to
little avail. That problem of a realistic interpretation of quantum mechanics
has become an open scandal \cite{vonkampan:scandal}. New so-called
``interpretations'' are created all the time
and no matter how bizarre, none are definitively abandoned in the
``demolition derby'' of interpretations. The
usual circular conversation in the philosophy of quantum mechanics [e.g.,
\cite{norsen:foundQM}; \cite{maudlin:qm}] typically considers the Copenhagen
interpretation associated with Niels Bohr \cite{faye:copenhagen}, together
with the realistic or ontic interpretations of Bohmian mechanics
\cite{durr:bohmian}, spontaneous localization \cite{gdw:spon-loc}, or
many-worlds \cite{wallace:many-worlds}. There is not even wide agreement on
what constitutes an ``interpretation'' or how
it should be constituted.

A new approach is needed. This paper focuses on giving the \textit{basic ideas} needed to understand (standard von Neumann-Dirac) QM, not on how they
connect together in the full-blown mathematics. That seems a reasonable place
to start.

\section{The basic idea of superposition}\label{sec2}
\subsection{Superposition as the flip-side of abstraction}\label{subsec2}

\begin{figure}[h]
	\centering
	\includegraphics[width=0.7\linewidth]{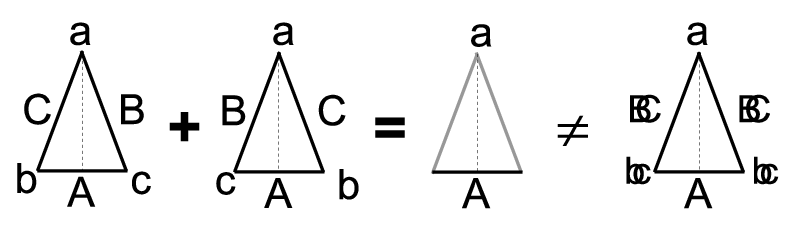}
	\caption{Superposition of two differently labeled isosceles triangles is
		indefinite where they differ}
	\label{fig:two-triangles}
\end{figure}

A glass half-full and a glass half-empty are the same thing viewed from
different perspectives. Abstraction and superposition have that relationship
(\cite{ell:abstr-super}; \cite{aerts:potentiality}). Given a set of entities
with some similarities and some differences, the process of abstraction
``abstracts away from the differences'' to
focus on the similarities. The ``abstraction\textquotedblright%
\ is definite on the similarities and indefinite on the differences. In Figure
\ref{fig:two-triangles}, we have two similar isosceles triangles with labeled edges and angles. The
process of abstraction to see only the similarities can also be seen as the
process of superposition by rendering the triangles indefinite on their
differences. This quantum superposition might be symbolized as: definite +
definite = indefinite, or, in more detail, superposition of similarities = definite on the one hand, and superposition of differences = indefinite on the other hand. The indefiniteness is not due to a lack of knowledge; it is objective indefiniteness \cite{ell:fop}. And as illustrated in Figure \ref{fig:two-triangles} on the right side, indefiniteness does not mean multiple definiteness as in the popular science descriptions of a particle going through both slits in the two-slit experiment.

\begin{figure}[h]
	\centering
	\includegraphics[width=0.7\linewidth]{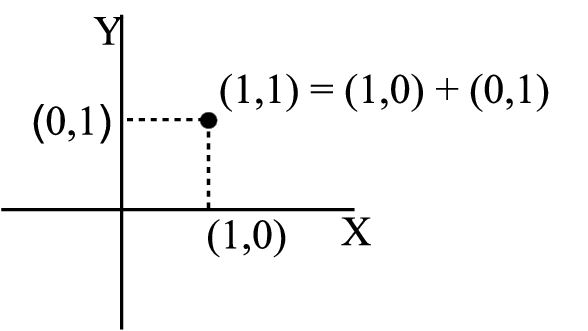}
	\caption{Superposition as indefinite between definte states}
	\label{fig:real-plane-superposition}
\end{figure}

This is also illustrated mathematically in the real plane in Figure \ref{fig:real-plane-superposition}. The states $(1,0)$ and $(0,1)$ are definite respectively in terms of ``x-ness'' and ``y-ness'' (with equal amplitudes). Their addition is the superposition $(1,1)$ that is indefinite between the two definite states of ``x-ness'' and ``y-ness.''

\begin{figure}[h]
	\centering
	\includegraphics[width=0.7\linewidth]{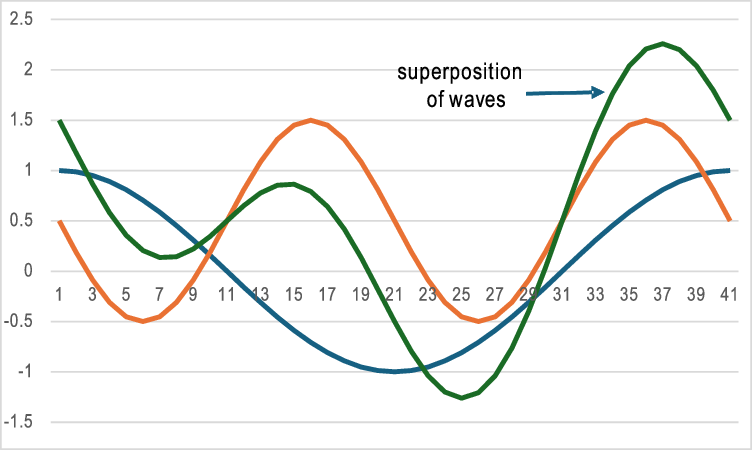}
	\caption{Classical superposition as definite + definite = definite}
	\label{fig:classical-superposition}
\end{figure}

A very common misunderstanding is to see quantum superposition (ontologically) as being like
the classical superposition of waves like water, sound, or electromagnetic
waves. The math is the same, but the interpretation should be different. The addition of two classical waves is just as definite or well-defined
as the summand waves--as illustrated in Figure \ref{fig:classical-superposition}. 

This is a key reinterpretation. That classical notion of
superposition is very different from superposition seen as the flip-side of
abstraction where the superposition is indefinite, e.g., indefinite between
going through the two-slits in the double-slit experiment or or going through the two arms in an interferometer \cite[p. 18]{auletta:qm}. 

\begin{quotation}
	\noindent Such analogies have led to the name ``Wave
	Mechanics'' being sometimes given to quantum mechanics. It is
	important to remember, however, that the superposition that occurs in quantum
	mechanics is of an essentially different nature from any occurring in the
	classical theory, as is shown by the fact that the quantum superposition
	principle demands indeterminacy in the results of observations in order to be
	capable of a sensible physical interpretation. The analogies are thus liable
	to be misleading. \cite[14]{dirac:principles}
\end{quotation}

The math is the same in the two
cases of classical and quantum superposition, i.e., vector addition, but the
interpretation in the two cases is very different. The complex numbers are the natural math to describe waves since the polar representation of a complex number is an amplitude and a phase. But that does not mean that the wave function is a physical wave. The reality that is described is that of superposition as indefiniteness in the states of quantum particles. R.I.G. Hughes referred to quantum indefiniteness as ``latency.''

\begin{quotation}
	\noindent The wave formalism offers a convenient mathematical representation of this latency, for not only can the mathematics of wave effects, like interference and diffraction, be expressed in terms of the addition of vectors (that is, their linear superposition; see \cite[chap. 29-5]{feynman:vol-1}, but the converse, also holds. \cite[303]{hughes:structure}
\end{quotation}

That is, the math of vector addition to describe quantum superposition as indefiniteness can always be seen in terms of classical wave effects such as interference and diffraction. The wave math of classical superposition thus \textit{tracks} (i.e., also describes) the quantum notion of superposition as indefiniteness. The math of classical wave motion (e.g., the ripple tank model of the two-slit experiment using classical water waves) \textit{also} describes the math of evolution of quantum particles in superposition or indefinite states (see Figure \ref{fig:full-case2-lattices} which illustrates this point in the context of the two-slit experiment). 

The huge payoff from interpreting quantum superposition as creating indistinctions, i.e., rendering differences as
indistinct (as in Figure \ref{fig:two-triangles}), is that we can see that the opposite
process of creating distinctions is state reduction (see later section), i.e.,
 
\begin{center}
	superposition = making indistinctions;
	
	state reduction = making distinctions.
\end{center}
With only the classical notion of superposition, state reduction (``collapse of the wave function'') appears as a mystery (the so-called ``measurement problem'') rather than just the inverse of quantum superposition.

\subsection{Superposition accounts for quantum amplitudes and the Born rule}\label{sec3}

\subsubsection{Superposition events in probability theory}\label{subsubsec2}

There is some agreement, starting at least with Paul Dirac \cite[Chapter
I]{dirac:principles}, that the idea of superposition is the key non-classical
notion in QM. Since the superposition of quantum states is another quantum
state, i.e., the linear combination of quantum states is a quantum state, the
idea of superposition is responsible for the quantum states forming a linear
vector space. Since our goal is conceptual understanding, not mathematical
generality, we stick to the finite-dimensional case. As Hermann Weyl put it
``no essential features of quantum mechanics are lost by using
the finite-dimensional model.'' \cite[257]{weyl:phil}

When a normalized quantum state vector is expressed in the basis of
eigenvectors (eigenstates) of an observable, then the Born rule states that
absolute square of the coefficients is the probability of an eigenstate being
the result of a measurement of that state by that observable. Steven Weinberg asked: ``Where does the Born rule come from?'' \cite[92]{weinberg:lectures} To answer that question, we might go back to
a suggestion of Gian-Carlo Rota; ``I will lay my cards on the
table: a revision of the notion of a sample space is my ultimate
concern.'' \cite[57]{rota:fubini}

\begin{quotation}
	\noindent Behind the Feynman integral there lurks an even more enticing (and
	even less rigorous) concept: that of an amplitude which is meant to be the
	quantum-mechanical analog of probability (one gets probabilities by taking the
	absolute values of amplitudes and squaring them: hence the slogan
	``quantum mechanics is the imaginary square root of
	probability theory\textquotedblright). A concept similar to that of a sample
	space should be brought into existence for amplitudes and quantum mechanics
	should be developed starting from this concept. \cite[229]{rota:indiscrete}
\end{quotation}

Hence we start with the ordinary notion in finite probability theory of a
\textit{sample space} $U$ with outcomes $u_{1},...,u_{n}$ which we initially
assume are equiprobable. An \textit{event }is a subset $S\subseteq U$ of
outcomes. The subset $S$ and the whole set $U$ have no structure connecting
the outcomes. The simplest possible idea for Rota's ``
revision'' is to postulate a new type of event, a
\textit{superposition event}, symbolized $\Sigma S$, where the outcomes are
``superposed'' instead of having unrelated
outcomes as in the classical discrete event $S$. And the simplest possible
assumption about the probabilities is to assume they are unchanged, i.e.,
$\Pr\left(  u_{i}|\Sigma S\right)  :=\Pr\left(  u_{i}|S\right)  $. At first
this looks like a bug but it is a feature since the same thing occurs in QM
where the probabilities for eigenstate outcomes in the superposition state are
the same as the probabilities in the corresponding complete decomposed mixed
state \cite[176]{auletta:qm}. Those two states can only be differentiated
by a measurement in another basis.

\subsubsection{Superposition and relation matrices}

How can $S$ and $\Sigma S$ be differently represented in the same basis? An
$n\times1$ column vector $\left\vert S\right\rangle $ of $0,1$-entries to
represent which outcomes are in the \textit{support} of the event $S$ or
$\Sigma S$, or even a similar vector of probabilities, will not do the job
since they are the same for both $S$ and $\Sigma S$.

One must move to an $n\times n$ matrix to represent the difference. We can
begin with the simplest case of $0,1$-matrices to represent the differences. A
\textit{binary relation} $R$ on $U$ is a subset $R\subseteq U\times U.$ The
relation can be represented by an $n\times n$ \textit{relation matrix }(also
called \textit{incidence matrix}) $\operatorname*{Rel}\left(  R\right)  $
where $\operatorname*{Rel}\left(  R\right)  _{jk}=1$ if $\left(  u_{j}%
,u_{k}\right)  \in R$, else $0$. Let $\Delta S\subseteq U\times U$ be the set
of self-pairs of elements in $S$, i.e., the diagonal $\Delta S=\left\{
\left(  u_{i},u_{i}\right)  |u_{i}\in S\right\}  $. Then the ordinary discrete
event $S$ is represented by the diagonal matrix $\operatorname*{Rel}\left(
\Delta S\right)  $ with the diagonal being the values of the characteristic
function $\chi_{S}:U\rightarrow\left\{  0,1\right\}  $, i.e., $\chi_{S}\left(
u_{i}\right)  =1$ if $u_{i}\in S$, else $0$. In contrast the superposition
event $\Sigma S$ is represented by $\operatorname*{Rel}\left(  S\times
S\right)  $ so that $\operatorname*{Rel}\left(  S\times S\right)  _{jk}=1$ if
$u_{j},u_{k}\in S$, else $0$. Intuitively, in the superposition event $\Sigma
S$, the elements of $S$ are rendered indefinite on their differences which has
been described as them being blurred, blobbed, smeared, or cohered together.
This cohering together of superposed outcomes is then represented by the
non-zero off-diagonal elements of $\operatorname*{Rel}\left(  S\times
S\right)  $. The difference between the two matrices $\operatorname*{Rel}%
\left(  \Delta S\right)  $ and $\operatorname*{Rel}\left(  S\times S\right)  $
is in those non-zero off-diagonal elements. Even though we are only dealing
(at first) with $0,1$-matrices, we can already see the foreshadowing of how
superposition is responsible for the non-classical aspects of QM.

\begin{quotation}
	\noindent For this reason, the off-diagonal terms of a density matrix ... are
	often called ``quantum coherences'' because
	they are responsible for the interference effects typical of quantum mechanics
	that are absent in classical dynamics. \cite[177]{auletta:qm}
\end{quotation}

\noindent The `waste case' of superposition is when $S$ is a singleton event
and accordingly in that case, $\operatorname*{Rel}\left(  \Delta S\right)
=\operatorname*{Rel}\left(  S\times S\right)  $ so we will only speak of a
``superposition event'' $\Sigma S$ when
$\left\vert S\right\vert \geq2$.

Two new phenomena appear in the matrix representation $\operatorname*{Rel}%
\left(  S\times S\right)  $ of a superposition event. One is the non-zero
off-diagonal entries representing the cohering together of the outcomes in the superposition event $\Sigma S$. The other new result is that only
$\operatorname*{Rel}\left(  S\times S\right)  $ can be obtained as the outer
(or external) product or ``square'' of the ''square root'' support $0,1$-vector $\left\vert S\right\rangle $
and its transpose $\left\vert S\right\rangle ^{t}$, i.e.,

\begin{center}
	$\left\vert S\right\rangle \left\vert S\right\rangle ^{t}=\operatorname*{Rel}%
	\left(  S\times S\right)  $.
\end{center}

\noindent If we represent the $j^{th}$ component of $\left\vert S\right\rangle
$ by $\left\langle u_{j}|S\right\rangle $, then $\operatorname*{Rel}\left(
S\times S\right)  _{jk}=\left\langle u_{j}|S\right\rangle \left\langle
u_{k}|S\right\rangle $. This is a \textit{very} important result since it
shows how, in an ultra-simple matrix representation of superposition, the ``square root''
vector $\left\vert S\right\rangle $ appears that foreshadows the vector of
quantum amplitudes in QM. And even the Born rule is foreshadowed in the fact
that: $\left\langle u_{i}|S\right\rangle ^{2}=\operatorname*{Rel}\left(
S\times S\right)  _{ii}$.

\subsubsection{Superposition and density matrices}

Dividing the relation matrices $\operatorname*{Rel}\left(  \Delta S\right)  $
and $\operatorname*{Rel}\left(  S\times S\right)  $ through by their trace
(sum of diagonal elements) turns them into \textit{density matrices
}\cite{weinberg:density} that represent quantum states:

\begin{center}
	$\rho\left(  \Sigma S\right)  =\frac{\operatorname*{Rel}(S\times
		S)}{\operatorname*{tr}\left[  \operatorname*{Rel}\left(  S\times S\right)
		\right]  }$ and $\rho\left(  \Delta S\right)  =\frac{\operatorname*{Rel}%
		\left(  \Delta S\right)  }{\operatorname*{tr}\left[  \operatorname*{Rel}%
		\left(  \Delta S\right)  \right]  }$.
\end{center}

\noindent They are density matrices for the equiprobable case obtained by
dividing through by $\operatorname*{tr}\left[  \operatorname*{Rel}\left(
S\times S\right)  \right]  =\operatorname*{tr}\left[  \operatorname*{Rel}%
\left(  \Delta S\right)  \right]  =\left\vert S\right\vert $ so that:

\begin{center}
	$\Pr\left(  u_{i}|\Sigma S\right)  :=\Pr\left(  u_{i}|S\right)  =\frac
	{1}{\left\vert S\right\vert }$ if $u_{i}\in S$, else $0$.
\end{center}

The density matrix $\rho\left(  \Sigma S\right)  $ is idempotent, i.e.,

\begin{center}
	$\rho\left(  \Sigma S\right)  ^{2}=\frac{1}{\left\vert S\right\vert
	}\operatorname*{Rel}\left(  S\times S\right)  \operatorname*{Rel}\left(
	S\times S\right)  \frac{1}{\left\vert S\right\vert }=\frac{1}{\left\vert
		S\right\vert }\left\vert S\right\rangle \left\vert S\right\rangle
	^{t}\left\vert S\right\rangle \left\vert S\right\rangle ^{t}\frac
	{1}{\left\vert S\right\vert }=\frac{1}{\left\vert S\right\vert }\left\vert
	S\right\rangle \left\vert S\right\vert \left\vert S\right\rangle ^{t}\frac
	{1}{\left\vert S\right\vert }=\frac{1}{\left\vert S\right\vert }%
	\operatorname*{Rel}\left(  S\times S\right)  =\rho\left(  \Sigma S\right)  $.
\end{center}

\noindent Hence it is a density matrix $\rho$ that is called a \textit{pure
	state} in QM, i.e., where $\rho^{2}=\rho$, while $\rho\left(  \Delta S\right)
$ ($\left\vert S\right\vert \geq2$) is what is called a \textit{mixed state. }

For example, suppose $U=\left\{  a,b,c\right\}  $ and $S=\left\{  a,c\right\}
$ with equiprobable outcomes. Then we have:

\begin{center}
	$\operatorname*{Rel}\left(  \Delta S\right)  =%
	\begin{bmatrix}
		1 & 0 & 0\\
		0 & 0 & 0\\
		0 & 0 & 1
	\end{bmatrix}
	$and $\operatorname*{Rel}\left(  S\times S\right)  =%
	\begin{bmatrix}
		1 & 0 & 1\\
		0 & 0 & 0\\
		1 & 0 & 1
	\end{bmatrix}
	$.
\end{center}

\noindent Dividing through by $\left\vert S\right\vert =2$ gives:

\begin{center}
	$\rho\left(  \Delta S\right)  =%
	\begin{bmatrix}
		\frac{1}{2} & 0 & 0\\
		0 & 0 & 0\\
		0 & 0 & \frac{1}{2}%
	\end{bmatrix}
	$ and $\rho\left(  \Sigma S\right)  =%
	\begin{bmatrix}
		\frac{1}{2} & 0 & \frac{1}{2}\\
		0 & 0 & 0\\
		\frac{1}{2} & 0 & \frac{1}{2}%
	\end{bmatrix}
	$
\end{center}

\noindent as well as:

\begin{center}
	$\rho\left(  \Sigma S\right)  ^{2}=%
	\begin{bmatrix}
		\frac{1}{2} & 0 & \frac{1}{2}\\
		0 & 0 & 0\\
		\frac{1}{2} & 0 & \frac{1}{2}%
	\end{bmatrix}
	^{2}=%
	\begin{bmatrix}
		\frac{1}{2} & 0 & \frac{1}{2}\\
		0 & 0 & 0\\
		\frac{1}{2} & 0 & \frac{1}{2}%
	\end{bmatrix}
	=\rho\left(  \Sigma S\right)  $.
\end{center}

The diagonal entries in any density matrix are non-negative reals that sum to
one like probabilities. The eigenvalues of a diagonal matrix like $\rho\left(
\Delta S\right)  $ are its diagonal entries. But the eigenvalues of a pure
state density matrix like $\rho\left(  \Sigma S\right)  $ are one with the
value of $1$ with the rest being zeros. Hence for any pure state $\rho
^{2}=\rho$ in QM, there is a normalized eigenvector $\left\vert s\right\rangle $
corresponding to the eigenvalue of $1$ so the spectral decomposition of $\rho$
as the sum of eigenvalues times the projectors to the eigenstates is:
$\rho=\left\vert s\right\rangle \left\langle s\right\vert $ so that $\rho$ is
again obtained as an \textit{outer} (or \textit{external}) \textit{product} or ``square'' of
a ``square root'' vector $\vert s \rangle $ times its (conjugate) transpose. In the case of our example, the
normalized eigenvector corresponding to the eigenvalue of $1$ for $\rho\left(
\Sigma S\right)  $ is (up to sign) $\left\vert s\right\rangle =\left[
\frac{1}{\sqrt{2}},0,\frac{1}{\sqrt{2}}\right]  ^{t}$ so that:

\begin{center}
	$\left\vert s\right\rangle \left\vert s\right\rangle ^{t}=%
	\begin{bmatrix}
		\frac{1}{\sqrt{2}}\\
		0\\
		\frac{1}{\sqrt{2}}%
	\end{bmatrix}
	\left[  \frac{1}{\sqrt{2}},0,\frac{1}{\sqrt{2}}\right]  =%
	\begin{bmatrix}
		\frac{1}{2} & 0 & \frac{1}{2}\\
		0 & 0 & 0\\
		\frac{1}{2} & 0 & \frac{1}{2}%
	\end{bmatrix}
	=\rho\left(  \Sigma S\right)  $.
\end{center}

The new item that appears for pure state density matrices $\rho$ is this state
vector $\left\vert s\right\rangle $ (corresponding to the eigenvalue of $1$)
such that $\rho=\left\vert s\right\rangle \left\vert s\right\rangle
^{t}=\left\vert s\right\rangle \left\langle s\right\vert $ (writing
$\left\vert s\right\rangle ^{t}$ as $\left\langle s\right\vert $ in the Dirac
notation). And, as in the example, the product of each entry in $\left\vert
s\right\rangle $ with corresponding entry in the (conjugate) transpose is the
diagonal entry in the density matrix, e.g., $\left\langle a|s\right\rangle
^{2}=\frac{1}{2}=\left\langle c|s\right\rangle ^{2}$ and $\left\langle
b|s\right\rangle ^{2}=0$ in the example, which in general is the Born rule.

Thus what we have derived starting with only the notion of a superposition
event $\Sigma S$ in an extended probability theory smoothly generalizes to the
general case in QM where transpose is the conjugate transpose and where the
square $\left\langle u_{i}|s\right\rangle ^{2}$ is the absolute value squared
$\left\vert \left\langle u_{i}|s\right\rangle \right\vert ^{2}$. The claim is
that this \textit{basic idea} of a superposition event $\Sigma S$ leads by this natural
logical progression to the notion of quantum amplitudes $\left\vert
s\right\rangle $ (or ``square roots'') and the Born rule $\Pr\left(  u_{i}|s\right)  =\left\vert
\left\langle u_{i}|s\right\rangle \right\vert ^{2}$.

Of course, in the context of the full mathematics of QM, there can be many
so-called ``derivations'' of the Born rule
\cite{vaidman:born}, not to mention the mathematics of the Gleason Theorem
\cite{gleason:born}. But such elegant and sophisticated results do not really
answer Weinberg's question: ``Where does the Born rule come
from?\textquotedblright. Our approach here is different, namely to give the
\textit{basic idea} behind the Born rule. And we have argued that the Born
rule and quantum amplitudes (whose absolute squares give the probabilities)
are natural consequences of just introducing the notion of a superposition
event into probability theory \cite{ell:born-again}.

There are a few other aspects that might be noted. The notion of
``support'' records only the information
about a scalar as to whether it is non-zero or zero. Given a state vector
$\left\vert \psi\right\rangle =\sum_{i=1}^{n}\alpha_{i}\left\vert
u_{i}\right\rangle $, the \textit{support set} is the set of basic vectors
with non-zero coefficients: $\operatorname*{supp}\left(  \left\vert
\psi\right\rangle \right)  =\left\{  \left\vert u_{i}\right\rangle |\alpha
_{i}\neq0\right\}  $, the \textit{support vector} $\left\vert
\operatorname*{supp}\left(  \left\vert \psi\right\rangle \right)
\right\rangle $ is the vector where the components $\left\langle u_{i}%
|\psi\right\rangle $ are replaced by their supports, i.e., $\langle
u_{i}\left\vert \operatorname*{supp}\left(  \left\vert \psi\right\rangle
\right)  \right\rangle =1$ if $\left\langle u_{i}|\psi\right\rangle \neq0$,
else $0$, and the \textit{support matrix} of a matrix replaces its entries by
their supports: $\operatorname*{supp}\left(  \rho\right)  _{jk}=1$ if
$\rho_{jk}\neq0$, else $0$. Then it is an easy result for any pure state
density matrix $\rho=\left\vert s\right\rangle \left\langle s\right\vert $ in
QM that:

\begin{center}
	$\operatorname*{supp}\left(  \rho\right)  =\operatorname*{Rel}\left(
	\operatorname*{supp}\left(  \left\vert s\right\rangle \right)  \times
	\operatorname*{supp}\left(  \left\vert s\right\rangle \right)  \right)  $.
\end{center}

\noindent which verifies our description of the superposition event $\Sigma S$ as $\operatorname*{Rel}(S\times S) $. In other words, the pattern of non-zero entries in any pure state
density matrix in QM is $S\times S$ for some subset $S$ of the basis set in which
the matrix is represented. This results follows from the fact that in any
(algebraic) field from $\mathbb{Z}_{2}$ to $\mathbb{C}$, the product of two scalars $\left\langle u_{j}|s\right\rangle $ and
$\left\langle u_{k}|s\right\rangle $ is non-zero if and only if (iff) both
scalars are non-zero.

A non-trivial question is the interpretation of the state vectors $\left\vert
s\right\rangle $ that give the corresponding pure state density matrices
$\rho=\left\vert s\right\rangle \left\langle s\right\vert $. Many
interpretations of QM take the state vectors, e.g., wave functions, as
ontological entities, rather than just a computational devices to compute the
probability amplitude and probabilities (via the Born rule) of possible measurement
outcomes. A classical discrete event and a superposition event are not ontological entities. They are simply a mathematical part of the extended probability theory \textit{to indicate the possible or potential outcomes}. We have derived the quantum amplitudes starting only with the notion
of a superposition event with no ontological assumptions. 

Moreover,
if the ``Schr\"{o}dinger wave'' is an ontic
wave, then it is very unclear why the absolute squares should be probabilities
rather than some notion of intensity. But our derivation shows how the ``square root'' state
vectors $\left\vert s\right\rangle $ arise out of adding superposition events
to \textit{probability theory} so the probabilistic interpretation is there from the beginning.

There is another argument, that might be mentioned, as to why the ``
Schr\"{o}dinger wave'' is not a description of an ontic wave. In William Rowan Hamilton's
optico-mechanical formulation of classical particle mechanics, the mathematics
of waves appears \cite[Sec. 7.9]{coopersmith:lazy}. There is even a classical
form of ``wave-particle duality.\textquotedblright%
\ ``Both optical and mechanical phenomena can be described in
wave terms as well as in particle terms.'' \cite[276]{lanczos:var-prin} Certainly no one interpreted this mathematical artifact
of waves as representing ontic waves \textit{in classical particle mechanics}. But
Schr\"{o}dinger introduced Planck's constant $h$ and reformulated the
Hamilton-Jacobi equation over the complex numbers to obtain his famous equation.

\begin{quotation}
	\noindent Schr\"{o}dinger had in 1927 the original idea of going beyond the
	analogy between geometrical optics and mechanics, established by Hamilton's
	partial differential equation, and changing over from the phase function
	$\phi$ to the wave function $\psi$. \cite[279]{lanczos:var-prin}
\end{quotation}

\noindent It is rather implausible to think, after these changes to obtain the
Schr\"{o}dinger equation, that the wave mathematics would suddenly describe
ontic waves instead of the indefinite superposition states of quantum particles. And since Hamilton's geometrical mechanics is the classical limit
as Planck's constant $h\rightarrow0$, how would an ontic wave mechanics turns into 
particle mechanics in the limit? In short, the wave functions of QM are about the indefinite superposition states of quantum particles as opposed to the fully definite states of classical particles; it is not about physical waves.

\section{The basic math of indefiniteness and definiteness}

\subsection{The logic of partitions (or equivalence relations)}

Abstracting away from the differences between to entities $u_{j}$ and $u_{k}$
means that they are equivalent in what characteristics that remain. In other
words, neglecting their differences means they are now in the same equivalence
class of some equivalence relation. And \textit{equivalence relation}
$E\subseteq U\times U$ is a reflexive, symmetric, and transitive binary
relation on $U$. Each element of $U$ has an \textit{equivalence class} of
other elements equivalent to it. Those equivalence classes are non-empty
disjoint subsets of $U$ that cover all of $U$ so they form a
\textit{partition} of $U$. Equivalence relations and partitions are
essentially the same concept but viewed from different perspectives, concepts
that Gian-Carlo Rota called ``cyrptomorphic\textquotedblright%
\ \cite[153]{kung;rota}. We will focus on the notion of a partition.

A \textit{partition} $\pi$ on $U=\left\{  u_{1},...,u_{n}\right\}  $ is a set
of nonempty disjoint subsets $\pi=\left\{  B_{j}\right\}  _{j=1}^{m}$ whose
union is all of $U$. A \textit{distinction} or \textit{dit} of $\pi$ is an
ordered pair $\left(  u_{j},u_{k}\right)  $ in different blocks of $\pi$ so
let $\operatorname*{dit}\left(  \pi\right)  \subseteq U\times U$ be the set of
distinctions of $\pi$. Then an \textit{indistinction} or \textit{indit} of
$\pi$ is an ordered pair of elements of $U$ in the same block of $\pi$ and
$\operatorname*{indit}\left(  \pi\right)  =U\times U-\operatorname*{dit}%
\left(  \pi\right)  $ is the equivalence relation associated with $\pi$.

The concepts of distinctions and indistinctions, distin-guishability and indistin-guishability, and definiteness and indefiniteness are the key concepts, the `natural language,' of quantum mechanics. Partition logic is their logic. They are responsible for the widespread belief that ``information,'' i.e., information-as-distinctions, plays a basic role in QM. The role of those concept is quite explicit in the Feynman rules.

\begin{quotation}
	If several alternative subprocesses, indistinguishable within the given physical arrangement, lead from the initial state to the final (registered) result, then the amplitudes for all the indistinguishable processes must be added to get the total amplitude for their combination (quantum law of superposition of amplitudes).
	
	If several distinguishable alternative processes lead from the initial preparation to the same final result, then the probabilities for all these processes must be added to get the total probability for the final result (law of addition of probabilities). \cite[110]{jaeger:qobjects}
\end{quotation}

\noindent This ontic role of distinctions and indistinctions, i.e., information, is quite unknown in the classical physics of fully definite particles. 

If $\sigma=\left\{  C_{j^{\prime}}\right\}  _{j^{\prime}=1}^{m^{\prime}}$ is
another partition on $U$, then $\sigma$ is \textit{refined} by $\pi$, denoted
$\sigma\precsim\pi$, if for any block $B_{j}\in\pi$ there is a block
$C_{j^{\prime}}\in\sigma$ containing it, i.e., $B_{j}\subseteq C_{j^{\prime}}$. The refinement partial ordering can also be expressed as just the inclusion
relation on ditsets:

\begin{center}
	$\sigma\precsim\pi$ iff $\operatorname*{dit}\left(  \sigma\right)
	\subseteq\operatorname*{dit}\left(  \pi\right)  $.
\end{center}

\noindent If $\Pi\left(  U\right)  $ is the set of partitions on $U$, then
refinement is a partial order on $\Pi\left(  U\right)  $. Moving upward in that partial order means making more distinctions. The top or maximal
partition of the partial order is the \textit{discrete partition}
$\mathbf{1}_{U}=\left\{  \left\{  u_{i}\right\}  \right\}  _{u_{i}\in U}$
where all the blocks are singletons. The bottom or minimal partition is the
\textit{indiscrete partition} $\mathbf{0}_{U}=\left\{  U\right\}  $ whose only
block is $U$. Since $U$ is the only block in $\mathbf{0}_{U}$, it has no
distinctions: $\operatorname*{dit}\left(  \mathbf{0}_{U}\right)  =\emptyset$.
The discrete partition makes all possible distinctions. Since no element can
be distinguished from itself, $\operatorname*{indit}\left(  \mathbf{1}%
_{U}\right)  =\Delta$ where diagonal $\Delta$ is $\Delta=\left\{  \left(
u_{i},u_{i}\right)  |u_{i}\in U\right\}  $ and $\operatorname*{dit}\left(
\mathbf{1}_{U}\right)  =U\times U-\Delta$. 

The \textit{join }of $\pi$ and $\sigma$, denoted $\pi\vee\sigma$, is the
partition whose blocks are all the nonempty intersections $B_{j}\cap
C_{j^{\prime}}$ of the blocks of $\pi$ and $\sigma.$ The join is the least
upper bound of $\pi$ and $\sigma$ in the refinement partial order, and its ditset is: $\operatorname*{dit}%
\left(  \pi\vee\sigma\right)  =\operatorname*{dit}\left(  \pi\right)
\cup\operatorname*{dit}\left(  \sigma\right)  $. Then by DeMorgan's law, its
indit set is: $\operatorname*{indit}\left(  \pi\vee\sigma\right)
=\operatorname*{indit}\left(  \pi\right)  \cap\operatorname*{indit}\left(
\sigma\right)  $, so the join of partitions corresponds to the intersection of
the corresponding equivalence relations. Since the intersection of two
equivalence relations is always an equivalence relation, the \textit{meet}
(greatest lower bound) of $\pi$ and $\sigma$, denoted $\pi\wedge\sigma$, is
the partition whose corresponding equivalence relation is the smallest
equivalence relation containing $\operatorname*{indit}\left(  \pi\right)
\cup\operatorname*{indit}\left(  \sigma\right)  $. That join and meet
operation on $\Pi\left(  U\right)  $ make it a \textit{lattice} which was
known in the 19th century (Dedekind and Schr\"{o}der).

\begin{figure}[h]
	\centering
	\includegraphics[width=0.7\linewidth]{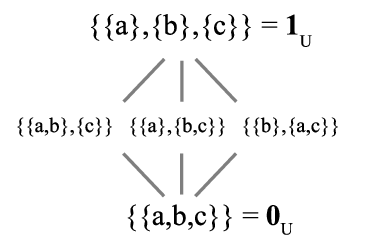}
	\caption{Partition lattice on $U=\left\{  a,b,c\right\}  $}
	\label{fig:three-partition-lattice}
\end{figure}

Figure \ref{fig:three-partition-lattice} gives the lattice of partitions for $U=\left\{  a,b,c\right\}  $
where refinement is indicated by the lines between partitions.

The ditset of the partition $\pi=\left\{  \left\{  a,b\right\}  ,\left\{
c\right\}  \right\}  $ is $\operatorname*{dit}\left(  \pi\right)  =\left\{
\left(  a,c\right)  ,\left(  b,c\right)  ,...\right\}  $ where the ellipsis
stands for the reversed ordered pairs. The ditset of the discrete partition
$\mathbf{1}_{U}$ is $\operatorname*{dit}\left(  \mathbf{1}_{U}\right)
=\left\{  \left(  a,b\right)  ,\left(  a,c\right)  ,\left(  b,c\right)
,...\right\}  $ so in moving up the refinement partial order from $\pi$ to
$\mathbf{1}_{U}$, means distinguishing $a$ from $b$ in block or equivalence
class $\left\{  a,b\right\}  $, i.e., making the distinctions $\left\{
\left(  a,b\right)  ,...\right\}  =\left\{  \left(  a,b\right)  ,\left(
b,a\right)  \right\}  =\operatorname*{dit}\left(  \mathbf{1}_{U}\right)
-\operatorname*{dit}\left(  \pi\right)  $.

Without at least the implication operation on partitions, the lattice of
partitions $\Pi\left(  U\right)  $ is not properly called a ``
logic.'' When the implication operation (and other logical
operations) were defined (in the 21st century), then the \textit{logic} of
partitions could be developed (\cite{ell:lop-book}; \cite{ell:partitions}).

\begin{figure}[h]
	\centering
	\includegraphics[width=0.7\linewidth]{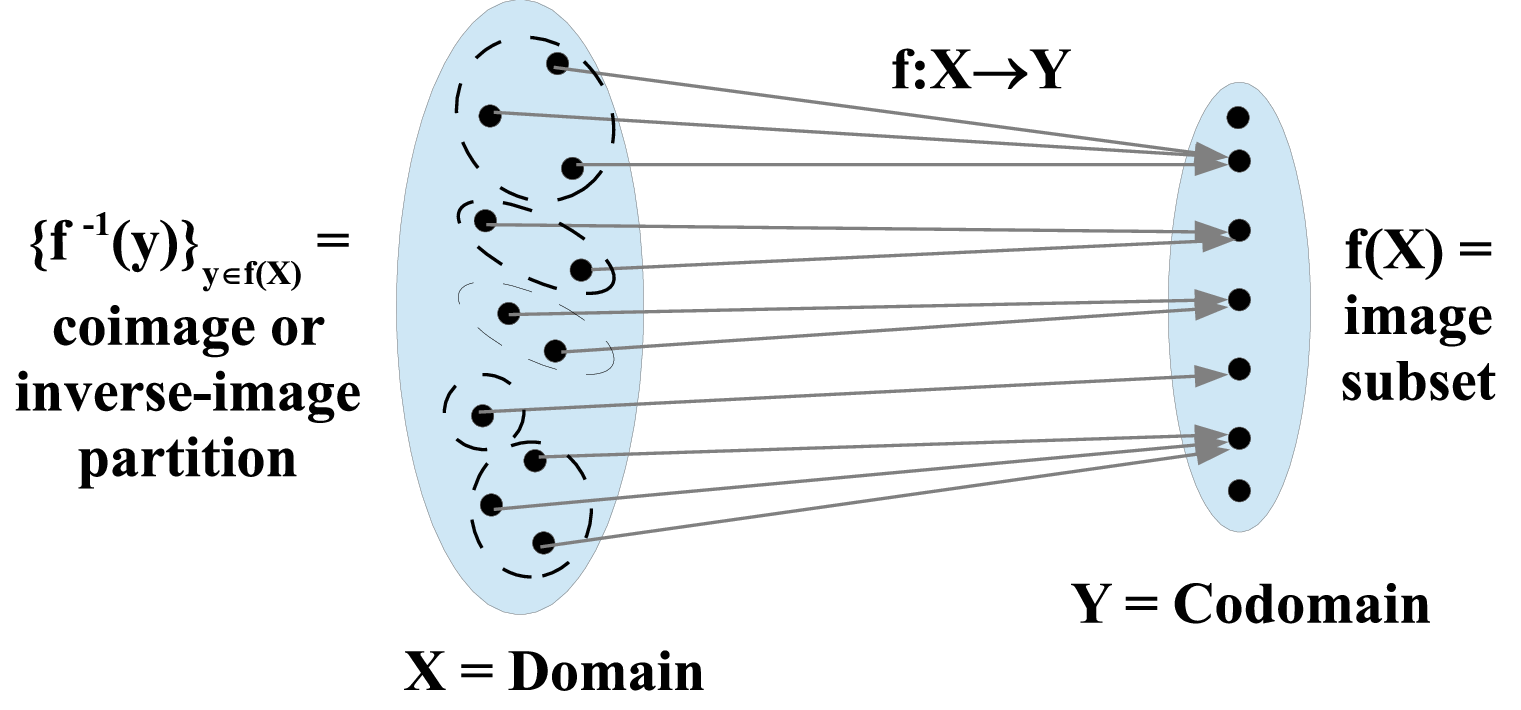}
	\caption{Duality between subsets and partitions illustrated with a function
		$f:X\rightarrow Y$}
	\label{fig:fcn-image-coimage}
\end{figure}

In category theory, there is the basic turn-around-arrows duality between
subsets (or subobjects or `parts') and partitions (or quotient objects).
``The dual notion (obtained by reversing the arrows) of `part'
is the notion of partition.'' \cite[85]{lawvere:sets} This
duality is illustrated in Figure \ref{fig:fcn-image-coimage} where the image of a set-function
$f:X\rightarrow Y$ is subset $f\left(  X\right)  \subseteq Y$ of the codomain
$Y$ and the coimage (or inverse-image) $f^{-1}=\left\{  f^{-1}\left(
y\right)  \right\}  _{y\in f\left(  X\right)  }$ is a partition on the domain
$X$.

Since the usual logic is the Boolean logic of subsets (usually presented in
the special case of the logic of propositions), the logic of partitions is, in
that sense, the dual logic to Boolean logic \cite{ell:partitions}. We will
eventually see that this basic duality is reflected in the difference between
classical mechanics and quantum mechanics \cite{ell:itsanddits}. This duality
comes out if we compare the Boolean lattice $\wp\left(  U\right)  $ and the
partition lattice $\Pi\left(  U\right)  $ by considering what happens in terms
of substance (or matter) and form \cite{aimsworth:FandM} moving from the
bottom to the top of each lattice.

\begin{figure}[H]
	\centering
	\includegraphics[width=0.9\linewidth]{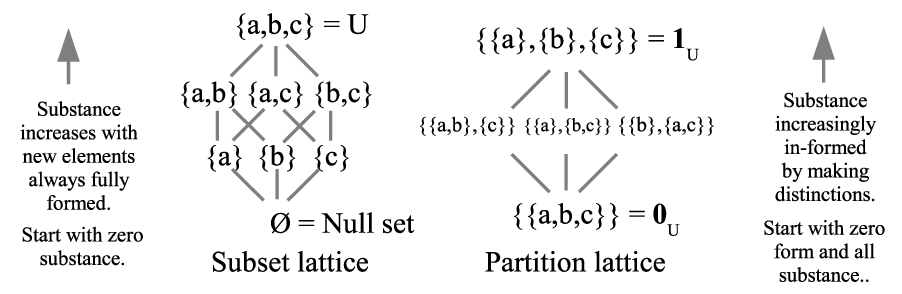}
	\caption{Duality illustrated between the lattice of subsets and the lattice
		of partitions}
	\label{fig:twocreationstories}
\end{figure}

The bottom of the Boolean lattice is the empty set $\emptyset$ and as we move
up the lattice, new substance is created in each subset until we reach the top
$U=\left\{  a,b,c\right\}  $. Each element in $U$ is always fully formed. In
the partition lattice, we see the opposite. At the bottom is the indiscrete
partition $\mathbf{0}_{U}=\left\{  U\right\}  =\left\{  \left\{
a,b,c\right\}  \right\}  $ so all the substance is there but in a totally
indefinite form with no distinctions. As we move up the lattice, there is no
new substance but the elements $a$, $b$, or $c$ become more in-formed with
distinctions between the elements of $U$. Elsewhere we have argued that
information is distinctions (\cite{ell:NF4IT}; \cite{ell:qic}). Hence moving
up the Boolean lattice means new substance is created but no new information
(since the elements are always fully distinguished). In the dual case, moving
up the partition lattice means new information-as-distinctions is created but
no new substance. That illustration of the duality is pictured in Figure \ref{fig:twocreationstories}.

\subsection{Mapping a quantum system into a partition lattice using support sets}

Think of $U$ as the basis set of orthonormal eigenvectors for some observable.
The support sets for quantum states can then be mapped into the partition
lattice $\Pi\left(  U\right)  $. The blocks in the partitions (equivalence
classes of equivalence relations) represent the support sets of quantum
states. In $\mathbf{0}_{U}$, there is only one block $U$ and it is the support
set of any (pure) quantum state that is a superposition of all the
eigenstates--like what we previously denoted as $\Sigma U$ (since all the
non-singleton subsets $S$ appearing in $\Pi\left(  U\right)  $ represent
superpositions previously denoted $\Sigma S$). At the other extreme is
$\mathbf{1}_{U}$ whose blocks are all the singletons $\left\{  u_{i}\right\}
$ for $u_{i}\in U$ so  $\mathbf{1}_{U}$ is the support set of the completely
decomposed mixed state--like what we previously described as the discrete
classical sample space $U$. 

In between are the support sets for all the mixed
states that could be obtained from a superposition of all the eigenstates
(modeled by $\mathbf{0}_{U}$) by the projective measurement or state reduction
by the observable, the operation which is mathematically described by the
L\"{u}ders mixture operation (\cite{luders:meas}; \cite{furry:luders};
\cite[279]{auletta:qm}). Those mixed states are special since the supports
of the states in the mixture are all disjoint, and that disjointness is
inherited from the disjointness of the eigenspaces of the observable via the
L\"{u}ders mixture operation representing projective measurement. And those disjoint supports cover all of $U$ since they result from projective measurement of $\mathbf{0}_U$, so those mixed state supports make up the partitions between the top and bottom of the partition lattice.

\begin{figure}[h]
	\centering
	\includegraphics[width=0.7\linewidth]{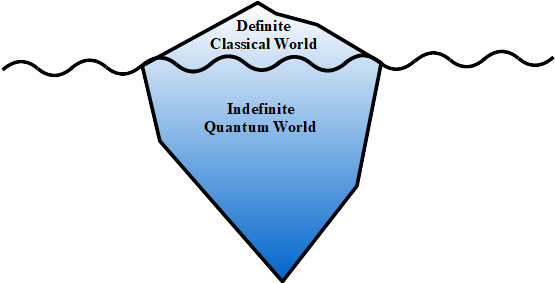}
	\caption{Iceberg metaphor: The definite classical world and the indefinite quantum world}
	\label{fig:iceberg-blue}
\end{figure}

The system can be represented
like an iceberg \cite[7]{kastner:alice}, the above-water part being the
classical mixed state $\mathbf{1}_{U}$, and the below-water part being all the
states involving at least one non-classical (i.e., indefinite) state, i.e., a superposition state as illustrated in Figure \ref{fig:iceberg-blue}

\begin{figure}[h]
	\centering
	\includegraphics[width=0.7\linewidth]{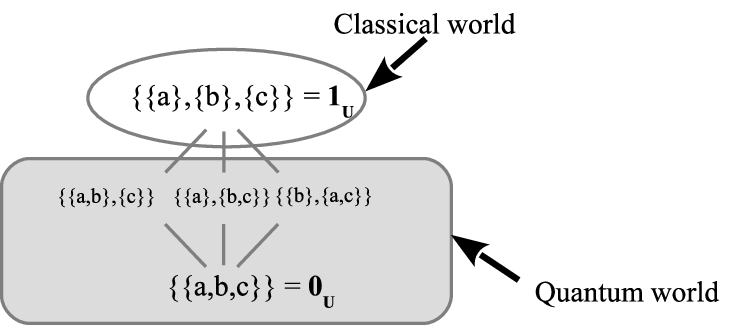}
	\caption{Partition lattice of possible support states of a quantum system of
		one particle with three eigenstates $a$, $b$, or $c$}
	\label{fig:iceberg-pix-3lattice}
\end{figure}

These two worlds can be represented mathematically the partition lattice.  The partition lattice $\Pi\left(  \left\{  a,b,c\right\}
\right)  $ (Figure \ref{fig:three-partition-lattice}) represents the supports of the possible states of the quantum system
representing one quantum particle in the three-dimensional Hilbert space $\mathbb{C}^{3}$. The striking thing is that the lattice of support sets boils the
possible quantum states of the particle down to show the clear separation
between the classical part $\mathbf{1}_{U}$ and all the other quantum states
represented by partitions with at least one non-singleton block representing
the support set of a superposition state. , as illustrated in Figure \ref{fig:iceberg-pix-3lattice}.

The iceberg picture and the partition lattice match up with the image of
reality divided into actuality (the classical part) and ``
potentiality'' (the indefinite quantum part) as advocated by Werner
Heisenberg \cite{heisen:phy-phil}, Abner Shimony \cite{shim:nat-world-II}, Gregg Jaeger \cite{jaeger:qobjects}, Diederik Aerts \cite{aerts:concept}, Ruth Kastner \cite{kastner:poss},
Leonardo Chiatti \cite{chiatti:poss}, and many others.

\begin{quotation}
	\noindent Heisenberg \cite[53]{heisen:phy-phil} used the term
	``potentiality'' to characterize a property
	which is objectively indefinite, whose value when actualized is a matter of
	objective chance, and which is assigned a definite probability by an algorithm
	presupposing a definite mathematical structure of states and properties.
	Potentiality is a modality that is somehow intermediate between actuality and
	mere logical possibility. That properties can have this modality, and that
	states of physical systems are characterized partially by the potentialities
	they determine and not just by the catalogue of properties to which they
	assign definite values, are profound discoveries about the world, rather than
	about human knowledge. \cite[6]{shim:vienna-yrbk}
\end{quotation}

It is particularly important to see that moving upward in the partition
lattice, e.g., from a quantum state (``potentiality'') to a classical state (actuality) means \textit{making distinctions} (since the refinement partial order on partitions is just inclusion of ditsets). This is the process of \textit{emergence} from the quantum world of indefiniteness into the classical world of definiteness, e.g., as illustrated in the partition lattice of Figure \ref{fig:twocreationstories}.

Any non-philosophical quantum theorist who believes a superposition state of an observable does
not have a definite value prior to measurement has already implicitly
acknowledged this quantum underworld of indefiniteness. The partition lattice
of indefinite support sets as partition blocks (below the top of classical
states) thus only adds some structure to what is commonly understood.

Heisenberg seems to have clothed his metaphysical speculations in discussions
of Greek philosophy and hence his use of Aristotle's notion of
``potentiality.'' But Shimony pointed out
that this was not a felicitous choice of concepts.

\begin{quotation}
	\noindent The historical reference should perhaps be dismissed, since quantum
	mechanical potentiality is completely devoid of teleological significance,
	which is central to Aristotle's conception. What it has in common with
	Aristotle's conception is the indefinite character of certain properties of
	the system. \cite[313-4]{shim:nat-world-II}
\end{quotation}

\noindent Indeed, the notion of indefiniteness is already a sufficient
description of the quantum underworld to contrast with the classical concepts
of full definiteness. Rather than ``actuality'' and ``potentiality'' (or ``
latency'' \cite{margenau:latency}; \cite{hughes:structure}),
reality divides into the quantum world of indefiniteness from which, with
distinctions, \textit{emerges} into the classical world of definiteness. The
admission of this quantum underworld of indefiniteness is the main ontological
implication of our analysis.

The partition lattice $\Pi\left(  U\right)  $ adds structure (simplified to the level of support sets) to the iceberg picture. This is well-illustrated in the mapping of the two von Neumann processes in the partition lattice for the two-slit experiment (see below). 

The emergence from indefinite to less indefinite or fully definite in the partition lattice matches up precisely with the case of state
reduction in the Feynman rules \cite[110]{jaeger:qobjects} due to
making distinctions, i.e., distinguishing between the superposed alternative
paths. Moving upward from an indefinite state to a less indefinite or even a
definite state, i.e., refinement, in the partition lattice represents a state
reduction--which in unnecessarily anthropocentric terms is often called a
``measurement'' even though there need be no
human involvement or interaction with a macroscopic apparatus. A human
measurement involves amplifying a quantum level state reduction to the level
of human observation but such technological considerations have no role in
quantum \textit{theory}.

\subsection{Join of partitions = partition lattice version of projective
	measurement}

Taking $U$ as an orthonormal eigenbasis for an observable (i.e., a Hermitian
or self-adjoint operator), the operator assigns a real number to each
eigenvector so it can be described as a numerical attribute $f:U\rightarrow\mathbb{R}$. The inverse-image $f^{-1}=\left\{  f^{-1}\left(  r\right)  \right\}  _{r\in
	f\left(  U\right)  }$ is a partition on $U$ and taking the state to be
measured as a partition $\pi$, then the result of the ``projection-valued measurement'' described by the L\"{u}ders
mixture operation is simply the join $f^{-1}\vee\pi$. The L\"{u}ders mixture
operation is defined in terms of projections and density matrices, but in the
basic environment of the powerset $\wp\left(  U\right)  $, a projection
operator $P_{S}$ defined by a subset $S\subseteq U$ takes any subset $T$ to
$S\cap T$, an idempotent operation. Hence the projection-valued measurement of
``mixed state'' $\pi=\left\{  B_{1},...,B_{m}\right\}  $ by the ``observable''
\ represented by $f^{-1}=\left\{  f^{-1}\left(  r\right)  \right\}  _{r\in
	f\left(  U\right)  }$ takes the blocks $B_{j}$ of $\pi$ to their projections
$f^{-1}\left(  r\right)  \cap B_{j}$, which are precisely the blocks of the join
$f^{-1}\vee\pi$. The indefinite states represented by the $B_{j}$'s are
reduced to the more definite states $f^{-1}\left(  r\right)  \cap B_{j}$ in
the mixture $f^{-1}\vee\pi$. The non-zero off-diagonal elements in
$\operatorname*{Rel}\left(  B_{j}\times B_{j}\right)  $ and $\rho\left(
\Sigma B_{j}\right)  $ represent the pairs of elements in the superposition
state $\Sigma B_{j}$ (hereafter $B_{j}$ means $\Sigma B_{j}$ unless indicated
otherwise) that are blobbed or cohered together like ``quantum
coherences'' \cite[177]{auletta:qm}. The L\"{u}ders
mixture operation is an operation that transforms a density matrix, such as

\begin{center}
	$\rho\left(  \pi\right)  =\sum_{B_{j}\in\pi}\Pr\left(  B_{j}\right)
	\rho\left(  B_{j}\right)  $
\end{center}

\noindent into the post-measurement density matrix $\widehat{\rho}(\pi)$. The
non-zero entries in $\rho\left(  \pi\right)  $ that are zeroed in that
operation $\rho(\pi)\rightarrow \widehat{\rho}(\pi)$, i.e., the pairs that are decohered in the join operation, are the
pairs with different $f$-values. That is the basic idea of projection-valued
measurement; it distinguishes superposed (i.e., indefinite) states that have
different eigenvalues.

\section{The pedagogical model of QM over (support) sets}

\subsection{The state space over $	\mathbb{Z}_{2}$}

A pedagogical (or `toy') model of QM, called \textit{QM/Sets}
(\cite{ell:qm/sets}; \cite{ell:simplified}) , can be constructed by working
with support vectors instead of the full state vectors over the complex
numbers $\mathbb{C}$. The support vectors have only $0,1$-components so they are vectors in the
vector space $\mathbb{Z}_{2}^{n}$ over the field $\mathbb{Z}_{2}=\left\{  0,1\right\}  $ where $1+1=0$. Of course, a lot of information is
lost about the non-zero coefficients in a state vector $\left\vert
\psi\right\rangle =\sum_{i=1}^{n}\alpha_{i}\left\vert u_{i}\right\rangle $,
but enough is retained to illustrate in a simple \textit{basic} manner some of
the paradoxical aspects of QM (see the treatment of the two-slit experiment below).

Since a $n$-ary $0,1$-vector in $\mathbb{Z}_{2}^{n}$ also defines a support subset $S\subseteq U$, the vectors can also treated
as subsets in the powerset $\wp\left(  U\right)  $. The addition of
two $0,1$-vectors in $\mathbb{Z}_{2}^{n}$ then corresponds to the addition of two subsets $S,T\in\wp\left(
U\right)  $ by the \textit{symmetric difference} operation, i.e.,

\begin{center}
	$S+T=\left(  S-T\right)  \cup\left(  T-S\right)  $.
\end{center}

\noindent For instance, if $U=\left\{  a,b,c\right\}  $, $S=\left\{
a,b\right\}  $, and $T=\left\{  b,c\right\}  $, then $S+T=\left\{
a,b\right\}  +\left\{  b,c\right\}  =\left\{  a,c\right\}  $ (since the $b$'s
cancel out). With that addition operation on $\wp\left(  U\right)  $, there is
a vector space isomorphism $\mathbb{Z}_{2}^{n}\cong\wp\left(  U\right)  $. As in any vector space, there are many
different basis sets, three of which are given in Table 1. Each row in the
table represents the same abstract vector or `ket' represented in a different
basis set. The left-most column gives the corresponding column vector in $\mathbb{Z}_{2}^{3}$ (where the superscript $t$ represents the transpose).

\begin{center}%
	\begin{tabular}
		[c]{|c|c|c|c|}\hline
		$\mathbb{Z}_{2}^{3}$ & $U=\left\{  a,b,c\right\}  $ & $U^{\prime}=\left\{  a^{\prime
		},b^{\prime},c^{\prime}\right\}  $ & $U^{\prime\prime}=\left\{  a^{\prime
			\prime},b^{\prime\prime},c^{\prime\prime}\right\}  $\\\hline\hline
		$\left[  1,1,1\right]  ^{t}$ & $\left\{  a,b,c\right\}  $ & $\left\{
		b^{\prime}\right\}  $ & $\left\{  a^{\prime\prime},b^{\prime\prime}%
		,c^{\prime\prime}\right\}  $\\\hline
		$\left[  1,1,0\right]  ^{t}$ & $\left\{  a,b\right\}  $ & $\left\{  a^{\prime
		}\right\}  $ & $\left\{  b^{\prime\prime}\right\}  $\\\hline
		$\left[  0,1,1\right]  ^{t}$ & $\left\{  b,c\right\}  $ & $\left\{  c^{\prime
		}\right\}  $ & $\left\{  b^{\prime\prime},c^{\prime\prime}\right\}  $\\\hline
		$\left[  1,0,1\right]  ^{t}$ & $\left\{  a,c\right\}  $ & $\left\{  a^{\prime
		},c^{\prime}\right\}  $ & $\left\{  c^{\prime\prime}\right\}  $\\\hline
		$\left[  1,0,0\right]  ^{t}$ & $\left\{  a\right\}  $ & $\left\{  b^{\prime
		},c^{\prime}\right\}  $ & $\left\{  a^{\prime\prime}\right\}  $\\\hline
		$\left[  0,1,0\right]  ^{t}$ & $\left\{  b\right\}  $ & $\left\{  a^{\prime
		},b^{\prime},c^{\prime}\right\}  $ & $\left\{  a^{\prime\prime},b^{\prime
			\prime}\right\}  $\\\hline
		$\left[  0,0,1\right]  ^{t}$ & $\left\{  c\right\}  $ & $\left\{  a^{\prime
		},b^{\prime}\right\}  $ & $\left\{  a^{\prime\prime},c^{\prime\prime}\right\}
		$\\\hline
		$\left[  0,0,0\right]  ^{t}$ & $\emptyset$ & $\emptyset$ & $\emptyset$\\\hline
	\end{tabular}

	Table 1: Ket table giving the isomorphisms $\mathbb{Z}_{2}^{3}\cong\wp\left(  U\right)  \cong\wp\left(  U^{\prime}\right)  \cong%
	\wp\left(  U^{\prime\prime}\right)  $
\end{center}

Taking $U$ as the computational basis, it is easy to see, for example, that
$\left\{  a^{\prime}\right\}  $, $\left\{  b^{\prime}\right\}  $, and
$\left\{  c^{\prime}\right\}  $ also form a basis, the $U^{\prime}$-basis, since:

$\left\{  b^{\prime}\right\}  +\left\{  c^{\prime}\right\}  =\left\{
b^{\prime},c^{\prime}\right\}  =\left\{  a,b,c\right\}  +\left\{  b,c\right\}
=\left\{  a\right\}  $,

$\left\{  a^{\prime}\right\}  +\left\{  b^{\prime}\right\}  +\left\{
c^{\prime}\right\}  =\left\{  a^{\prime},b^{\prime},c^{\prime}\right\}
=\left\{  a,b\right\}  +\left\{  a,b,c\right\}  +\left\{  b,c\right\}
=\left\{  b\right\}  $, and

$\left\{  a^{\prime}\right\}  +\left\{  b^{\prime}\right\}  =\left\{
a^{\prime},b^{\prime}\right\}  =\left\{  a,b\right\}  +\left\{  a,b,c\right\}
=\left\{  c\right\}  $.

\subsection{von Neumann's two quantum processes}

John von Neumann postulated two and only two types of quantum processes: Type
I are the state reductions and Type II are the evolutions according to the
Schr\"{o}dinger equation \cite{vonn:mfqm}. What is the basic idea? We have
already seen that the basic idea in Type I state reduction is a process of
making distinctions. Hence the natural definition of the other Type II
processes would be processes that don't make distinctions. The distinctness of
two (normalized) quantum states $\psi$ and $\phi$ is their inner product
$\left\langle \psi|\phi\right\rangle $. If the inner product is zero, they are
totally distinct with no `overlap.' If the inner product is one, then there is
total overlap, i.e., they are the same state. Hence the basic idea of a Type
II process is one that doesn't make distinctions so the measure of the
indistinctness of distinctness of two quantum states, i.e., the inner product,
is preserved--which is a unitary transformation. The connection to the
solutions to the Schr\"{o}dinger equation is given by Stone's theorem
\cite{stone:thm}.

Another way to characterize a unitary transformation is that it transforms
orthonormal basis sets into orthonormal basis sets. There are no inner
products in vector spaces over finite fields like $\mathbb{Z}_{2}$ in our pedagogical model QM/Sets. Hence the corresponding idea in a
finite vector space is a transformation that is \textit{non-singular}, i.e.,
transforms basis sets into basis sets so that is the Type II process assumed
in the model.

\subsection{Probabilities in QM/Sets}

The Dirac brackets in QM give the ``overlap'' between two states where the minimal
overlap is $\left\langle \psi|\phi\right\rangle =0$ for states that are
orthogonal (or disjoint) and maximal overlap is $\left\langle \psi
|\phi\right\rangle =1$ when they are the same state. In QM/Sets, there is an
obvious notion of overlap, the cardinality of the intersection, that takes its
values in the natural numbers $\mathbb{N}$. That is, for $S,T\in\wp\left(  U\right)  $:

\begin{center}
	$\left\langle S|_{U}T\right\rangle :=\left\vert S\cap T\right\vert $.
\end{center}

The ket $\left\vert T\right\rangle $ denotes the ket of $T\in\wp\left(
U\right)  $ and is in that sense basis-independent, but the `bra'
$\left\langle S\right\vert _{U}$ must be taken as basis-dependent as indicated
by the subscript $U$ since the intersection $S\cap T$ requires that both $S$
and $T$ be subsets of $U$.

The unitary transformations in QM are replaced by the non-singular
transformations in the vector space $\wp\left(  U\right)  $ that carry a basis
set to a basis set. Accordingly, that preserves the overlaps instead of the
inner products. For instance in the non-singular transformation from the
$U$-basis to the $U^{\prime}$-basis, i.e., $\left\{  a\right\}
\rightsquigarrow\left\{  a^{\prime}\right\}  $, $\left\{  b\right\}
\rightsquigarrow\left\{  b^{\prime}\right\}  $, and $\left\{  c\right\}
\rightsquigarrow\left\{  c^{\prime}\right\}  $, we have:

\begin{center}
	$\left\langle S|_{U}T\right\rangle =\left\langle S^{\prime}|_{U^{\prime}%
	}T^{\prime}\right\rangle $.
	
	Preservation of bra-kets under non-singular transformations
\end{center}

A projection operator $P$ on a vector space is an operator that is idempotent,
i.e., $PP=P$. For the universe set $U$ with the disjoint basis $\left\{
\left\{  u_{i}\right\}  \right\}  _{u_{i}\in U}$, the projection operator
$\left\{  u_{i}\right\}  \cap():\wp\left(  U\right)  \rightarrow\wp\left(
U\right)  $ takes $S$ to $\left\{  u_{i}\right\}  \cap S$ which is $\left\{
u_{i}\right\}  $ if $u_{i}\in S$ and $\emptyset$ otherwise. The characteristic
function $\chi_{\left\{  u_{i}\right\}  }:U\rightarrow\mathbb{Z}_{2}$, with
value $1$ at $u_{i}$, else $0$, defined the same projection operator $\widehat{\chi}_{\left\{
	u_{i}\right\}  }=\left\{  u_{i}\right\}  \cap():\wp\left(  U\right)
\rightarrow\wp\left(  U\right)  $ by $\widehat{\chi}_{\left\{  u_{i}\right\}
}\left\{  u_{j}\right\}  =\chi_{\left\{  u_{i}\right\}  }\left(  u_{j}\right)
\left\{  u_{i}\right\}  $. Then the sum of these projection operators over the
whole $U$-basis is the identity operator:

\begin{center}
	$\sum_{u_{i}\in U}\left\{  u_{i}\right\}  \cap()=\sum_{u_{i}\in U}\widehat{\chi
	}_{\left\{  u_{i}\right\}  }=I():\wp\left(  U\right)  \rightarrow\wp\left(
	U\right)  $.
\end{center}

In QM, given an orthonormal (ON) basis $\left\{  \left\vert u_{i}%
\right\rangle \right\}  _{i=1}^{n}$ of the Hilbert space $V$, the
characteristic function $\chi_{\left\{  u_{i}\right\}  }:U\rightarrow\left[
0,1\right]  $ defines the projection operator $\widehat{\chi}_{\left\{
	u_{i}\right\}  }=\left\vert u_{i}\right\rangle \left\langle u_{i}\right\vert
:V\rightarrow V$ and the sum of the ket-bra projection operators is also the
identity operator:

\begin{center}
	$\sum_{i=1}^{n}\left\vert u_{i}\right\rangle \left\langle u_{i}\right\vert
	=\sum_{i=1}^{n}\widehat{\chi}_{\left\{  u_{i}\right\}  }=I:V\rightarrow V$
	
	Completeness of the ket-bra sum
\end{center}

\noindent Now $\left\langle \left\{  u_{i}\right\}  |_{U}S\right\rangle
=\left\langle S|_{U}\left\{  u_{i}\right\}  \right\rangle =\left\vert
S\cap\left\{  u_{i}\right\}  \right\vert =\chi_{S}\left(  u_{i}\right)  $. Then any bra-ket $\left\langle
S|_{U}T\right\rangle $ can be resolved using the ket-bra sum:

\begin{center}
	$\sum_{u_{i}\in U}\left\langle S|_{U}\left\{  u_{i}\right\}  \right\rangle
	\left\langle \left\{  u_{i}\right\}  |_{U}T\right\rangle =\sum_{u_{i}\in
		U}\chi_{S}\left(  u_{i}\right)  \chi_{T}\left(  u_{i}\right)  =\left\vert
	S\cap T\right\vert =\left\langle S|_{U}T\right\rangle $
\end{center}

\noindent which is the QM/Sets version of the QM:

\begin{center}
	$\left\langle \psi|\varphi\right\rangle =\sum_{i}\left\langle \psi
	|u_{i}\right\rangle \left\langle u_{i}|\varphi\right\rangle $
	
	Resolution of unity by ket-bra sum
\end{center}

In QM, the magnitude or norm of a vector $\psi$ is often denoted $\left\vert
\psi\right\vert =\sqrt{\left\langle \psi|\psi\right\rangle }$ but that
conflicts with our notation $\left\vert S\right\vert $ for cardinality, so we
will use $\left\Vert \psi\right\Vert =\sqrt{\left\langle \psi|\psi
	\right\rangle }$ for the norm in QM and the corresponding norm in QM/Sets is:

\begin{center}
	$\left\Vert S\right\Vert _{U}=\sqrt{\left\langle S|_{U}S\right\rangle }%
	=\sqrt{\left\vert S\right\vert }$
	
	Norm in QM/Sets
\end{center}

\noindent which takes values in the real numbers $\mathbb{R} $. Applied to the resolution of unity:

\begin{center}
	$\left\Vert S\right\Vert _{U}^{2}=\left\langle S|_{U}S\right\rangle
	=\sum_{u\in U}\left\langle S|_{U}\left\{  u_{i}\right\}  \right\rangle
	\left\langle \left\{  u_{i}\right\}  |_{U}S\right\rangle =\left\vert
	S\right\vert $
\end{center}

\noindent which in QM is:

\begin{center}
	$\left\Vert \psi\right\Vert ^{2}=\left\langle \psi|\psi\right\rangle =\sum
	_{i}\left\langle \psi|u_{i}\right\rangle \left\langle u_{i}|\psi\right\rangle
	=\sum_{i}\left\langle u_{i}|\psi\right\rangle ^{\ast}\left\langle u_{i}%
	|\psi\right\rangle $
\end{center}

\noindent where $\left\langle u_{i}|\psi\right\rangle ^{\ast}=\left\langle
\psi|u_{i}\right\rangle $ is the complex conjugate of $\left\langle u_{i}%
|\psi\right\rangle $.

Since the non-zero amplitudes are replaced by ones in the move to support vectors, the outcomes are assumed equiprobable in QM/Sets. In QM, a vector can be normalized at any time, but in QM/Sets, normalization
is only done when probabilities are computed, so to better draw out the
analogies, we will not necessarily assume a vector $\psi$ is normalized. When
a state $\psi$ is measured in the measurement basis $\left\{
\left\vert u_{i}\right\rangle \right\}  $, then the probability of obtaining
$u_{i}$ is given by the Born Rule:

\begin{center}
	$\Pr\left(  u_{i}|\psi\right)  =\frac{\left\Vert \left\langle u_{i}%
		|\psi\right\rangle \right\Vert ^{2}}{\left\Vert \psi\right\Vert ^{2}}$
\end{center}

\noindent and the corresponding Born Rule formula in QM/Sets is:

\begin{center}
	$\Pr\left(  u_{i}|_{U}S\right)  =\frac{\left\Vert \left\langle \left\{
		u_{i}\right\}  |_{U}S\right\rangle \right\Vert _{U}^{2}}{\left\Vert
		S\right\Vert _{U}^{2}}=\frac{\left\vert \left\{  u_{i}\right\}  \cap
		S\right\vert }{\left\vert S\right\vert }=\left\{
	\begin{array}
		[c]{c}%
		1/\left\vert S\right\vert \text{ if }u_{i}\in S\\
		0\text{ if }u_{i}\notin S
	\end{array}
	\right.  $.
\end{center}

\noindent And given a numerical attribute $f:U\rightarrow \mathbb{R}$, then $%
f^{-1}\left( r\right) \cap ():\wp \left( U\right) \rightarrow \wp \left(
U\right) $ is a projection operator and the probability of getting $r\in f\left( U\right) $ when measuring $S$ is:
\begin{center}
	$\Pr (r|S)=\frac{\left\Vert f^{-1}\left( r\right) \cap S\right\Vert ^{2}}{%
		\left\Vert S\right\Vert }=\frac{|f^{-1}\left( r\right) \cap S|}{\left\vert
		S\right\vert }$.
\end{center}

\noindent In QM, if $P_{r}$ is the projection operator to the eigenspace of the
eigenvalue $r$, then the probability of getting that eigenvalue when measuring $\vert\psi \rangle$ is:

\begin{center}
	$\Pr \left( r|\vert \psi \rangle \right) =\frac{\left\Vert P_{r}\left\vert \psi
		\right\rangle \right\Vert ^{2} }{\left\Vert \left\vert \psi \right\rangle
		\right\Vert ^{2} }$.
\end{center}

In this manner, QM/Sets produces a simplified version of QM over support sets. The crucial application is the two-slit experiment.

\subsection{The two-slit experiment: the setup}

The two-slit experiment is the best example to illustrate the basic ideas of
QM. Indeed, according to Richard Feynman, it contains ``the
only mystery'' \cite[Sec. 1.1]{feynman:v3mill-ed}. The
pedagogical model greatly simplifies the model but reproduces
``the only mystery'' in the form of the
question: ``With no detection at the slits, how does the
particle get from the two-slit screen to the detection wall without going
through one of the slits--in which case there would be no interference
effects?''. The mystery is often covered up with a bit of
magic or legerdemain called ``wave-particle complementarity.'' With no detection at the slits, the
particle suddenly turns into a wave which, in a certain sense,
``goes through both slits'' like in the
classroom ripple-tank demonstration \cite{llowarch:rippletank} using classical waves. But a wave cannot register
at just one point on the detection wall so the wave thoughtfully turns back
into a particle after the interference effects. The actual explanation can be
easily seen in the pedagogical model without any such magic.

\begin{figure}[h]
	\centering
	\includegraphics[width=0.7\linewidth]{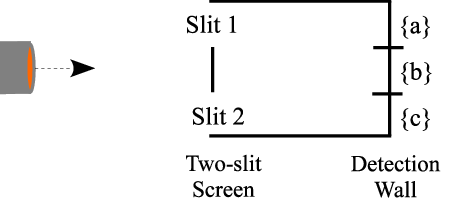}
	\caption{Setup for the two-slit experiment}
	\label{fig:two-slit-setup}
\end{figure}

The assumed discrete dynamics is that of the non-singular transformation where
in each time period, the $U$-basis turns into the $U^{\prime}$-basis, i.e.,
$\left\{  a\right\}  \rightsquigarrow\left\{  a^{\prime}\right\}  =\left\{
a,b\right\}  $, $\left\{  b\right\}  \rightsquigarrow\left\{  b^{\prime
}\right\}  =\left\{  a,b,c\right\}  $, and $\left\{  c\right\}
\rightsquigarrow\left\{  c^{\prime}\right\}  =\left\{  b,c\right\}  $. The
three states $\left\{  a,b,c\right\}  $ of the one particle system represent
vertical distance positions with the particle emitter at $\left\{  b\right\}
$ and the two slits on the screen at $\left\{  a\right\}  $ and $\left\{
c\right\}  $ as illustrated in Figure \ref{fig:two-slit-setup}.

In the first time period, the particle moves from the particle emitter at
$\left\{  b\right\}  $ to the screen at $\left\{  b^{\prime}\right\}
=\left\{  a,b,c\right\}  $. Then the first state reduction takes place where
the particle either hits the screen at $\left\{  b\right\}  $ with probability
$\Pr\left(  \left\{  b\right\}  |_{U}\left\{  a,b,c\right\}  \right)
=\frac{\left\vert \left\{  b\right\}  \cap U\right\vert }{\left\vert
	U\right\vert }=\frac{1}{3}$. Otherwise, the particle arrives at the screen in
the superposition state $\left\{  a,c\right\}  $ with probability $\Pr\left(
\left\{  a,c\right\}  |_{U}\left\{  a.b.c\right\}  \right)  =\frac{\left\vert
	\left\{  a,c\right\}  \cap\left\{  U\right\}  \right\vert }{\left\vert
	U\right\vert }=\frac{2}{3}$. Then we have the two cases of Case 1 of detectors
at the slits or Case 2 of no detectors at the slits.

\subsection{Case 1: Detection at the slits}

The detectors \textit{distinguish} between the two states in superposition at
the screen: $\left\{  a,c\right\}  =\left\{  \text{Going through slit 1, Going
	through slit 2}\right\}  $ so the superposition is reduced to one of the
components. If it is reduced to $\left\{  a\right\}  =\left\{  \text{Going
	through slit 1}\right\}  ,$ then it evolves to the superposition $\left\{
a,b\right\}  $ at the detection wall which \textit{distinguishes} between the
two components so the superposition $\left\{  a^{\prime}\right\}  =\left\{
a,b\right\}  $ reduces to $\left\{  a\right\}  $ or to $\left\{  b\right\}  $
with probability $\frac{1}{2}$ each. Similarly, if the $\left\{  a,c\right\}
$ superposition at the screen reduces to $\left\{  c\right\}  =\left\{
\text{Going through slit 2}\right\}  $, then it evolves to the superposition
$\left\{  c^{\prime}\right\}  =\left\{  b,c\right\}  $ at the detection wall.
And then the wall \textit{distinguishes} between those two components so they
reduce to $\left\{  b\right\}  $ or $\left\{  c\right\}  $ with $\frac{1}{2}$
probability each. The probabilities multiply along the paths (Feynman Rule 3.5 law of multiplication of probabilities \cite[111]{jaeger:qobjects}), so we have at $\left\{  a\right\}  $ and $\left\{  c\right\}  $ on the wall:

\begin{center}
	$\Pr\left(  a|\text{wall}\right)  =\frac{2}{3}\frac{1}{2}\frac{1}{2}=\frac
	{1}{6}=\Pr\left(  c|\text{wall}\right)  $.
\end{center}

\noindent There are two paths for the particle to reach $\left\{  b\right\}  $
at the wall, so (Feynman rule 3.3 law of addition of probabilities \cite[110]{jaeger:qobjects}) those probabilities add so that:

\begin{center}
	$\Pr\left(  b|\text{wall}\right)  =\frac{2}{3}\frac{1}{2}\frac{1}{2}+\frac
	{2}{3}\frac{1}{2}\frac{1}{2}=\frac{1}{6}+\frac{1}{6}=\frac{1}{3}$.
\end{center}

\begin{figure}[h]
	\centering
	\includegraphics[width=0.7\linewidth]{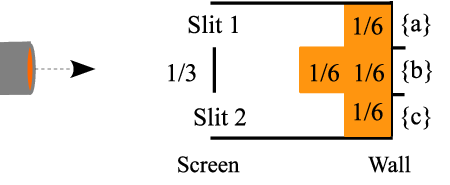}
	\caption{Probabilities of the particle (starting at the emitter) of hitting
		the detection wall}
	\label{fig:case1-probs}
\end{figure}

\noindent The bar graph of the Case 1 probabilities is illustrated in Figure \ref{fig:case1-probs}.

\subsection{Case 2: No detection at the slits}

In Case 2, there is no state reduction in the superposition $\left\{
a,c\right\}  $ at the screen so that superposition state, which is
below-the-water in the iceberg picture, i.e., in the quantum world, continues
to evolve according to the dynamics:

\begin{center}
	At the screen $\left\{  a,c\right\}  =\left\{  a\right\}  +\left\{  c\right\}
	\rightsquigarrow\left\{  a^{\prime}\right\}  +\left\{  c^{\prime}\right\}
	=\left\{  a,b\right\}  +\left\{  b,c\right\}  =\left\{  a,c\right\}  $ at the wall.
\end{center}

\noindent Then the detection wall \textit{distinguishes} between the
components of $\left\{  a,c\right\}  $ at the wall so $\left\{  a\right\}  $
or $\left\{  c\right\}  $ occur with $\frac{1}{2}$ probability each. Then the
probabilities at the wall are:

\begin{center}
	$\Pr\left(  a|\text{wall}\right)  =\frac{2}{3}\frac{1}{2}=\frac{1}{3}%
	=\Pr\left(  c|\text{wall}\right)  $ and $\Pr\left(  b|\text{wall}\right)  =0$.
\end{center}

\begin{figure}[h]
	\centering
	\includegraphics[width=0.7\linewidth]{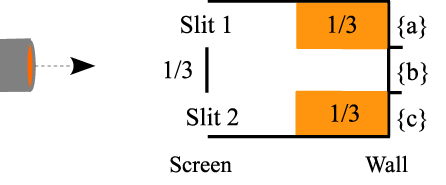}
	\caption{Probabilities of the particle (starting at the emitter) of hitting
		at the detection wall}
	\label{fig:case2-probs}
\end{figure}

\noindent The bar graph of the Case 2 probabilities is illustrated in Figure \ref{fig:case2-probs}.

\noindent Immediately we see the striped pattern due to the interference--in
this case the destructive interference at ${b}$ in the evolution:

\begin{center}
	At the screen $\left\{  a,c\right\}  \rightsquigarrow\left\{  a,b\right\}
	+\left\{  b,c\right\}  =\left\{  a,c\right\}  $ at the wall.
\end{center}

\subsection{Analysis of the two-slit experiment}

This is the analysis in the simplified pedagogical model of QM/Sets that
illustrates the basic ideas involved in the two-slit experiment. To complete
the explanation, we need to bring in the iceberg/partition-lattice picture to illustrate the two
cases using the partition lattices. At the screen, the two states of $\left\{
a\right\}  =\left\{  \text{Going through slit 1}\right\}  $ and $\left\{
c\right\}  =\left\{  \text{Going through slit 2}\right\}  $ are
\textit{classical} above-the-water states, neither of which occurs in Case 2
of no detection at the slits. As Feynman put it: ``We must
conclude that when both holes are open it is \textit{not true} that the
particle goes through one hole or the other.'' \cite[536]{feynman: prob-in-qm} Hence the question: ``In Case 2,
which slit does the particle go through?'' falsely assumes
that one of those two classical events occurs. Classical reality emerges from the indefinite quantum world by the making of distinctions. But in Case 2, the quantum particle does not emerge to the classical states of going through one slit or the other at the two-slit screen.

\begin{figure}[h]
	\centering
	\includegraphics[width=0.7\linewidth]{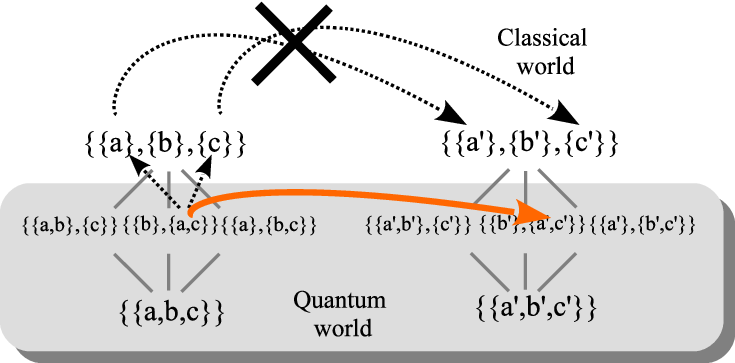}
	\caption{Case 2 evolution of the particle in the superposition state
		$\left\{  a,c\right\}  $ which is in the quantum world}
	\label{fig:iceberg-two-slit}
\end{figure}

This is an example of trying to
fit quantum (below-the-water) events into a classical (above-the-water) framework of thinking so
it appears to be a mystery how the particle can get to the detection wall
without one of the classical states of going through one of the slits
occurring. The dead end of the reasoning using classical states is what prompts the magic of ``particle-wave complementarity'' to picture the quantum particle as suddenly turning into a classical wave ``going through both slits.'' But that is not what happens. Since we interpreted superposition non-classically as indefiniteness and since the particle retains the indefinite state $\{a,c\}$ in Case 2, the particle evolves in the
below-the-water quantum world of indefiniteness as illustrated in Figure \ref{fig:iceberg-two-slit}.

The partition lattice is a super-simplified picture of reality which consists
of the classical reality of fully definite states and the quantum world of
indefinite superposition mixed and pure states.

	\begin{enumerate}
		\item Since von Neumann's Type I state reductions result from making
		distinctions, they are always represented by upward arrows (from indefinite to
		more definite states) in the partition lattice.
		
		\item The von Neumann Type II evolutions are represented by the non-upward arrow (horizontal or downward) arrows in the partition lattice.
	\end{enumerate}

\begin{figure}[h]
	\centering
	\includegraphics[width=0.7\linewidth]{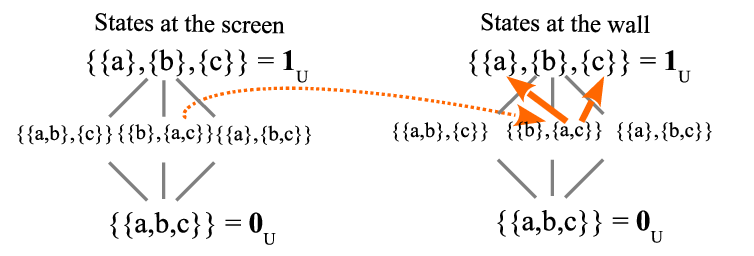}
	\caption{Two von Neumann processes in Case 2}
	\label{fig:full-case2-lattices}
\end{figure}

Figure \ref{fig:full-case2-lattices} illustrates both the (under-water or quantum world) evolution
(dotted horizontal arrows) from screen to the wall in Case 2 and then the
state reductions (upward solid arrows) at the wall.

\begin{figure}[h]
	\centering
	\includegraphics[width=0.7\linewidth]{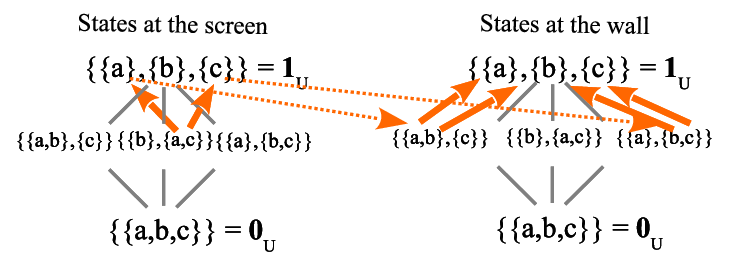}
	\caption{Two von Neumann processes in Case 1}
	\label{fig:full-case1-lattices}
\end{figure}

The partition lattices can also be used to represent the state reductions and
evolutions in Case 1. Again in Figure \ref{fig:full-case1-lattices}, the dotted (non-upward) arrows are
Type II evolutions and the solid upward arrows are state reductions.

\begin{figure}[h]
	\centering
	\includegraphics[width=0.7\linewidth]{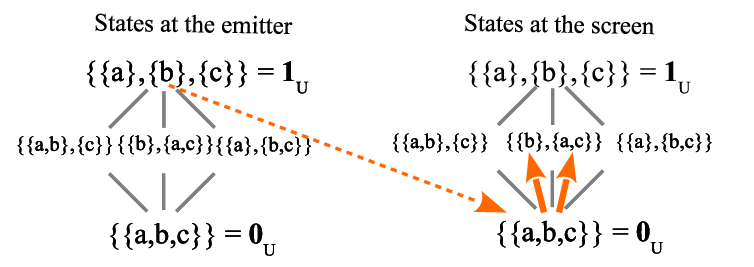}
	\caption{Evolution from emitter to screen and state reduction at the screen}
	\label{fig:emitter2screen}
\end{figure}

In both Case 1 and Case 2, there was the same evolution from the emitter to the screen and the same state reduction at the screen as illustrated in Figure \ref{fig:emitter2screen}.

\section{Linearization: Two-way street between sets and vector spaces}

\subsection{The Yoga of Linearization}

There is an extensive two-way connection between set concepts and vector space
concepts. It gives a translation dictionary between sets and vector spaces. In
enumerative combinatorics, set concepts are correlated with the corresponding
concepts in finite vector spaces of order $q=p^{n}$ (where $p$ is a prime) so
the vector space concept is called the ``$q$%
-analog'' \cite{goldman-rota:IV}. But we are concerned with
the dictionary relating set concepts to the corresponding vector space
concepts in the Hilbert spaces of QM so we will call them ``QM-analogs." This is an important tool to illustrate the basic ideas of QM by
distilling them down to the corresponding (support) set concepts as we have seen with
QM/Sets and the analysis of the two-slit experiment.

\begin{center}%
	\begin{tabular}
		[c]{|c|}\hline
		Yoga of Linearization\\\hline\hline
		Given a basis set of a vector space, apply the set concept to the basis set\\
		and then what is linearly generated\\ is the corresponding or QM-analog vector
		space concept.\\\hline
	\end{tabular}
	
\end{center}

We start by taking $U$ to be a basis set of a finite-dimensional Hilbert space
$V$, e.g., an orthonormal eigenbasis for an observable. Then the set notion of
a subset $S\subseteq U$ linearly generates a subspace $\left[  S\right]
\subseteq V$. The cardinality $\left\vert S\right\vert $ of $S$ equals the
dimension $\dim\left(  \left[  S\right]  \right)  $ of $\left[  S\right]  $. A
\textit{real-valued numerical attribute on }$U$ is a function $f:U\rightarrow\mathbb{R}$. This set concept generates a Hermitian operator $\widehat{f}:V\rightarrow V$ by
the equation $\widehat{f}u_{i}=f\left(  u_{i}\right)  u_{i}$ (where we ignore the
difference between set element $u_{i}$ and basis vector $\left\vert u_{i}\right\rangle $). When
appropriate, the QM-analog is indicated by the $\widehat{hat}$ on the set
version as in $\widehat{f}$ and $f$. Each value $r$ of $f:U\rightarrow\mathbb{R}$ is an eigenvalue of $\widehat{f}$, each subset $S$ $\subseteq f^{-1}(r)$ is a
constant set of $f$ so the set version of the eigenvector-eigenvalue equation
$\widehat{f}u_{i}=ru_{i}$ is $f\upharpoonright S=rS$, which means $f$ restricted to $S$
is a constant set of $f$ with value $r$. 

Each block $B_{j}$ in a set partition $\pi$
on $U$ generates a subspace $\left[  B_{j}\right]  $ and the set of those
subspaces $\widehat{\pi}=\left\{  \left[  B_{j}\right]  \right\}  _{B_{j}\in\pi}$
is a \textit{direct sum decomposition} (DSD) of $V$ as we would expect from
the $q$-analog: ``Direct-sum decompositions are a $q$-analog
of partitions of a finite set.'' \cite[764]%
{bender-goldman:genfcns}

A set partition $\pi=\left\{  B_{1},...,B_{m}\right\}  $ on $U$ could have
been defined as a set of non-empty subsets so that each non-empty subset
$S\subseteq U$ is uniquely expressed as the union of subsets of the $B_{j}$'s,
i.e., $S=\cup_{B_{j}\cap S\neq\emptyset}B_{j}\cap S$. Similarly, a DSD can be
defined as a set of non-trivial subspaces $\left\{  V_{j}\right\}  _{j=1}^{m}$
of $V$ such that every non-zero vector in $V$ is uniquely expressed as the sum
of vectors from the $V_{j}$'s. When the partition on $U$ is $f^{-1}=\left\{
f^{-1}\left(  r\right)  \right\}  _{r\in f\left(  U\right)  }$, then the
QM-analog DSD corresponding to $f^{-1}$ is the DSD $\widehat{f^{-1}}=\left\{
\left[  f^{-1}\left(  r\right)  \right]  \right\}  _{r\in f\left(  U\right)
}$ of eigenspaces of the Hermitian operator $\widehat{f}:V\rightarrow V$. If $f$
is just an attribute defined by a characteristic function $\chi_{S}%
:U\rightarrow\left\{  0,1\right\}  $, then the QM-analog operator $\widehat{\chi
}_{\left[  S\right]  }u_{i}=\chi_{S}\left(  u_{i}\right)  u_{i}$ is the
projection operator projecting to the space $\left[  S\right]  $. The spectral
decomposition of $\widehat{f}$ is:

\begin{center}
	$\widehat{f}=\sum_{r\in f\left(  U\right)  }r\widehat{\chi}_{\left[  f^{-1}\left(
		r\right)  \right]  }  $.
\end{center}

\noindent Then we can work backwards to see that the set-version of the
spectral decomposition for the numerical attribute $f$ is obtained by
``taking off the (operator) hats'':

\begin{center}
	$f=\sum_{r\in f\left(  U\right)  }r\chi_{f^{-1}(r)}$.
\end{center}

An important example of the correlations starts with a number of numerical
attributes $f,g,...,h:U\rightarrow\mathbb{R}$ all defined on the same basis set $U$. Then the QM-analog Hermitian
operators $\widehat{f},\widehat{g},...,\widehat{h}$ are all \textit{commuting} operators
(since $U$ provides a common basis of simultaneous eigenvectors). Moreover, if
the join $f^{-1}\vee g^{-1}\vee...\vee h^{-1}=\mathbf{1}_{U}$, i.e., has all
blocks of cardinality one, then each element $u_{i}\in U$ is uniquely
specified by the ordered-tuple of attribute values $\left(  f\left(
u_{i}\right)  ,g\left(  u_{i}\right)  ,...,h\left(  u_{i}\right)  \right)  $.
We might say that numerical attributes defined on the same set are
\textit{compatible} and if their join is the discrete partition $\mathbf{1}%
_{U}$, then they are a Complete Set of Compatible Attributes, a CSCA.

Similarly, each inverse-image partition $f^{-1}=\left\{  f^{-1}\left(
r\right)  \right\}  _{r\in f\left(  U\right)  }$ defines the QM-analog DSD
$\widehat{f^{-1}}=\left\{  \left[  f^{-1}\left(  r\right)  \right]  \right\}
_{r\in f\left(  U\right)  }$. The \textit{join of} \textit{DSDs of commuting
	operators} $\widehat{f^{-1}}\vee\widehat{g^{-1}}\vee...\vee\widehat{h^{-1}}$
is defined as the DSD of non-zero subspaces obtained by the intersections of
the eigenspaces of the operators. If the join $\widehat{f^{-1}}\vee
\widehat{g^{-1}}\vee...\vee\widehat{h^{-1}}$ of those DSDs is the DSD $\widehat{\mathbf{1}_{U}}$ of
subspaces of cardinality one or rays (instead of blocks of cardinality one),
then the set of operators is said to be \textit{complete}, i.e., a Complete
Set of Commuting Operators, a CSCO \cite[57]{dirac:principles}. And, as in
the set case, each eigenvector $u_{i}$ in the basis set of simultaneous
eigenvectors $U$ is then uniquely characterized by the ordered-tuple of eigenvalues
$\left(  f\left(  u_{i}\right)  ,g\left(  u_{i}\right)  ,...,h\left(
u_{i}\right)  \right)  $. That set version of a CSCA shows the basic idea
behind Dirac's CSCOs.

These correlations (using $\widehat{hat}$ notation) are summarized in Table 2.

\begin{center}%
	\begin{tabular}
		[c]{|c|c|}\hline
		Set math & QM-analogs\\\hline\hline
		Subset $S\subseteq U=\left\{  u_{1},...,u_{n}\right\}  $ & {\small Subspace
		}$\left[  S\right]  \subseteq V$\\\hline
		Numerical attribute $f:U\rightarrow		\mathbb{R}$ & Hermitian op. $\widehat{f}u_{i}=f\left(  u_{i}\right)  u_{i}$\\\hline
		Constant set $\ f\upharpoonright S=rS$, $r\in f\left(  U\right)  $ &
		Eigenvector $\widehat{f}v=rv$\\\hline
		Value $r$ of $f$ & Eigenvalue $r$ of $\widehat{f}$\\\hline
		Set of constant $r$-sets $\wp\left(  f^{-1}\left(  r\right)  \right)  $ &
		Eigenspace of $r$, $\left[  f^{-1}\left(  r\right)  \right]  $\\\hline
		Partition: $f^{-1}=\left\{  f^{-1}\left(  r\right)  \right\}  _{r\in f\left(
			U\right)  }$ & {\small Add} $\widehat{hat}$ {\small for DSD:} $\widehat{f^{-1}%
		}=\left\{  \left[  f^{-1}\left(  r\right)  \right]  \right\}  _{r\in f\left(
			U\right)  }$\\\hline
		$\chi$-function $\chi_{f^{-1}\left(  r\right)  }:U\rightarrow\left\{
		0,1\right\}  $ & {\small Projection op.} $\widehat{\chi}_{f^{-1}\left(  r\right)
		}u_{i}=\chi_{f^{-1}\left(  r\right)  }\left(  u_{i}\right)  u_{i}$\\\hline
		{\small Spectral decomp.} $f=\sum_{r\in f\left(  U\right)  }%
		r\chi_{f^{-1}\left(  r\right)  }$ & Add $\widehat{hats}$: $\widehat{f}=\sum_{r\in
			f\left(  U\right)  }r\widehat{\chi}_{f^{-1}\left(  r\right)  }$\\\hline
		{\small Num. attrib.} $f,g,...,h:U\rightarrow		\mathbb{R}$ {\small same} $U$ & {\small Add} $\widehat{hats}$ {\small for commuting
			op.:} $\widehat{f},\widehat{g},...,\widehat{h}$\\\hline
		$U=$ {\small same domain of }$f,g,...,h$ & $U=$ {\small Simultaneous
			eigenvectors} $\widehat{f},\widehat{g},...,\widehat{h}$\\\hline $f^{-1}\vee g^{-1}$ $=\left\{  f^{-1}\left(  r\right)  \cap g^{-1}\left(
		s\right)  \neq\emptyset\right\}  $ & $\widehat{f^{-1}}\vee\widehat{g^{-1}%
		}=\left\{  \left[  f^{-1}\left(  r\right)  \right]  \cap\left[  g^{-1}\left(
		s\right)  \right]  \neq\left\{  0\right\}  \right\}  $\\\hline
		$f^{-1}\vee g^{-1}\vee...\vee h^{-1}=\mathbf{1}_{U}$ & $\widehat{f^{-1}}\vee\widehat{g^{-1}}\vee
		...\vee\widehat{h^{-1}}=\widehat{\mathbf{1}_{U}}$\\
		$u_{i}\leftrightarrow\left(  f\left(  u_{i}\right)  ,g\left(  u_{i}\right)
		,...,h\left(  u_{i}\right)  \right)  $ & $u_{i}\leftrightarrow\left(  f\left(
		u_{i}\right)  ,g\left(  u_{i}\right)  ,...,h\left(  u_{i}\right)  \right)  $\\
		Set version CSCA & Dirac's CSCO\\\hline
	\end{tabular}

	Table 2: Correlations between set concepts and QM-analog concepts
\end{center}

\subsection{Non-commutativity}

In the early days of QM, the non-commutativity of observables seemed like a
key characteristic of QM as opposed to classical mechanics, e.g., Dirac's
q-numbers (linear operators) versus the classical c-numbers \cite{dirac:dev}.
But this seems to put the emphasis in the wrong place. After all, the
non-commutativity of matrix multiplication is a feature of vector spaces
\textit{per se} so the emphasis should be put on the quantum states forming a
vector space in the first place--which puts the emphasis back on superposition
rather than non-commutativity.

The vector space version of a set partition is a direct sum decomposition or
DSD. What is usually taken as ``quantum
logic'' is the logic of (closed) subspaces of a Hilbert space
\cite{birk-vN:logic}. Since subspaces are the vector space version of subsets,
that quantum logic of subspaces is the quantum version of the Boolean logic of
subsets. Since partitions are category-theoretically dual to subsets and since
DSDs are the vector space version of partitions, there is another dual \textit{quantum logic of direct sum decompositions} \cite{ell:DSDs}, which could also be viewed
as the quantum logic of observables (since the observable differs from its DSD
of eigenspaces by including the eigenvalues associated with the
eigenspaces--just as a numerical attribute $f:U\rightarrow\mathbb{R}$ differs from the partition $\left\{  f^{-1}\left(  r\right)  \right\}
_{r\in f\left(  U\right)  }$ only by including the numbers assigned to the blocks).

In forming the join of two partitions on the same $U$, we take the blocks of
the join to be the non-empty intersections of the blocks from the partitions.
But given two partitions on different universe sets $U$ and $U^{\prime}$, the
intersection is completely undefined. But that changes in the vector space
version of partitions, DSDs. A DSD is a basis-free notion. It is a set of
subspaces and subspaces always have intersections as subspaces. That is the
basic idea that creates the new possibilities of commuting DSDs, non-commuting
DSDs, and even conjugate DSDs.

Let $\left\{  V_{j}\right\}  _{j=1}^{m}$ and $\left\{  W_{j^{\prime}}\right\}
_{j^{\prime}=1}^{m^{\prime}}$ be two DSDs of a vector space $V$. We take the
set of non-zero intersections $V_{j}\cap W_{j^{\prime}}\neq\left\{  0\right\}
$ in a join-like operation. If the DSDs were the DSDs of eigenspace DSDs of
two observables $F,G:V\rightarrow V$, then the non-zero vectors in the blocks
would be simultaneous eigenvectors for the two observables. But the big
difference is that the vectors in those blocks need not span the whole space.
The join-like operation is only a ``proto-join.'' 

Let $\mathcal{SE}$ be the vector space spanned by the vectors in those
non-zero intersections $V_{j}\cap W_{j^{\prime}}$. If those two DSDs were the
eigenspace DSDs of two observable operators $F$ and $G$, then it is a theorem
\cite[68]{ell:poiqr} that: $\mathcal{SE}$ is the kernel (i.e., subspace of elements mapped to zero) of the commutator
$\left[  F,G\right]  =FG-GF$ of the operators. Now $F$ and $G$ \textit{commute
}if the commutator is the zero operator, i.e., if its kernel $\mathcal{SE}$ is
the whole space $V$ in which case the proto-join of the two DSDs is a proper
join, a join we have already seen in the analysis of Dirac's CSCOs. If
$\mathcal{SE}$ $\neq V$, then the two DSDs are \textit{non-commuting} and if
$\mathcal{SE=}\left\{  0\right\}  $, then the DSDs are \textit{conjugate}. In the
transition from an observable operator to its eigenspace DSD, the information
that is lost is the distinct eigenvalues associated with eigenspaces. But that
information is irrelevant for the definitions of commuting, not commuting, and
conjugate so those are, as we have seen, properties of the DSDs. 

This distinction between linear operators and DSDs is more pronounced when we move
to other vector spaces. In the spaces over $\mathbb{Z}_{2}$, the only linear operators are projection operators but far more general
are the DSDs $\left\{  \left[  f^{-1}\left(  r\right)  \right]  \right\}
_{r\in f\left(  U\right)  }$ in $\mathbb{Z}_{2}^{\left\vert U\right\vert }$ resulting from numerical attributes
$f:U\rightarrow\mathbb{R}$.

\textbf{Commutativity}. To illustrate this analysis, let $U=\left\{  u_{1},...,u_{n}\right\}  $ in
QM/Sets. Then any two numerical attributes $f,g:U\rightarrow\mathbb{R}$ defined on that same $U$ with have inverse-image partitions $f^{-1}=\left\{
f^{-1}\left(  r\right)  \right\}  _{r\in f\left(  U\right)  }$ and
$g^{-1}=\left\{  g^{-1}\left(  s\right)  \right\}  _{s\in g\left(  U\right)
}$ on $U$ and the blocks of the two partitions will generate two DSDs on $\mathbb{Z}_{2}^{n}\cong\wp\left(  U\right)  $. The subspaces in the join-like operation
on those two DSDs will be the subspaces generated by the blocks of $f^{-1}\vee
g^{-1}$ which contain all basis elements $\left\{  u_{1}\right\}
,...,\left\{  u_{n}\right\}  $ of $\wp\left(  U\right)  $ so the proto-join of
the DSDs spans the whole space and is thus a join of DSDs. That is an example
of commuting DSDs.

\textbf{Non-commutativity}. To consider an example of non-commutativity, consider the $U$-basis and the
$U^{\prime}$-basis of $\mathbb{Z}_{2}^{3}$ given in Table 1 and used in the two-slit example. Consider the
numerical attribute $f:U=\left\{  a,b,c\right\}  \rightarrow\mathbb{R}$ of $f\left(  a\right)  =1$, $f\left(  b\right)  =2$, and $f\left(  c\right)
=3$. On the $U^{\prime}$-basis, consider the numerical attribute $g:U^{\prime
}=\left\{  a^{\prime},b^{\prime},c^{\prime}\right\}  \rightarrow\mathbb{R}$ where $g\left(  a^{\prime}\right)  =1=g\left(  b^{\prime}\right)  $ and
$g\left(  c^{\prime}\right)  =2$. Then the DSD determined by $f$ is the set of
three subspaces: $\left\{  \left\{  \emptyset,\left\{  a\right\}  \right\}
,\left\{  \emptyset,\left\{  b\right\}  \right\}  ,\left\{  \emptyset,\left\{
c\right\}  \right\}  \right\}  $. The DSD determined by $g$ is the set of two
subspaces: $\left\{  \left\{  \emptyset,\left\{  a^{\prime}\right\}  ,\left\{
b^{\prime}\right\}  ,\left\{  a^{\prime},b^{\prime}\right\}  \right\}
,\left\{  \emptyset,\left\{  c^{\prime}\right\}  \right\}  \right\}  $. We may
then express the $g$-DSD in the computational $U$-basis as: $\left\{  \left\{
\emptyset,\left\{  a,b\right\}  ,\left\{  a,b,c\right\}  ,\left\{  c\right\}
\right\}  ,\left\{  \emptyset,\left\{  b,c\right\}  \right\}  \right\}  $.
Then when we take the join-like operation or proto-join by taking all the
intersections of subspaces, then many are just the zero space such as
$\left\{  \emptyset,\left\{  a\right\}  \right\}  \cap\left\{  \emptyset
,\left\{  a,b\right\}  ,\left\{  a,b,c\right\}  ,\left\{  c\right\}  \right\}
=\left\{  \emptyset\right\}  $, but only one intersection is non-trivial:
$\left\{  \emptyset,\left\{  c\right\}  \right\}  \cap\left\{  \emptyset
,\left\{  a,b\right\}  ,\left\{  a,b,c\right\}  ,\left\{  c\right\}  \right\}
=\left\{  \emptyset,\left\{  c\right\}  \right\}  =\mathcal{SE}$. However, the
vectors in this subspace hardly span the whole space so those two DSDs are
non-commuting but not conjugate.

\textbf{Conjugacy}. The vector spaces $\mathbb{Z}_{2}^{2m}$ for $m>1$ have a special structure much like the Fourier
transformation between conjugate variables in the full math of QM. The
simplest such space for $m=2$ is $\mathbb{Z}_{2}^{4}\cong\wp\left(  U\right)  $ for $U=\left\{  a,b,c,d\right\}  $. From
the $U$-basis, we canonically construct the $\hat{U}$-basis for $\hat
{U}=\left\{  \hat{a},\hat{b},\hat{c},\hat{d}\right\}  $ where the
circumflex operation (unrelated to our previous $\widehat{hat}$ notation) just
leaves out the element under the circumflex. Thus $\left\{  \hat
{a}\right\}  =\left\{  b,c,d\right\}  $, $\left\{  \hat{b}\right\}  =\left\{
a,c,d\right\}  $, $\left\{  \hat{c}\right\}  =\left\{  a,b,c\right\}  $. and
$\left\{  \hat{d}\right\}  =\left\{  a,b,c\right\}  $. And, as in the Fourier
transformation, the reverse operation on the $\hat{U}$-basis gives back the
$U$-basis. Thus $\left\{  \hat{b},\hat{c},\hat{d}\right\}  =$ $\left\{
a,c,d\right\}  +\left\{  a,b,d\right\}  +\left\{  a,b,c\right\}  =\left\{
a\right\}  $ since all the elements but $a$ occurs an even number of times so
they cancel out in the addition $\operatorname{mod}(2)$ while $a$ occurs an
odd number of times. And it works similarly for the other elements, e.g.,
$\left\{  b\right\}  =\left\{  \hat{a},\hat{c},\hat{d}\right\}  $ and so
forth. For $\mathbb{Z}_{2}^{2m}$ of even dimension, the circumflex-vectors form a basis but not for vector
spaces over $\mathbb{Z}_{2}$ of odd dimension. And for $U=\left\{  a,b\right\}  $, $\left\{  \hat
{a}\right\}  =\left\{  b\right\}  $ and $\left\{  \hat{b}\right\}  =\left\{
a\right\}  $, so $\hat{U}$-basis in that case is the same as the $U$-basis. That is why the Fourier-like transform is for vector spaces over $\mathbb{Z}_{2}$ of even dimension greater than two.

In QM, the Fourier transformation gives conjugate bases which gives the
conjugacy between the quantum variables such as position and momentum
\cite[Sec. 4.1.2]{deharo-butterfield:duality}. In our example of $\mathbb{Z}_{2}^{2m}$ ($m>1$), any $U$-basis has a conjugate $\hat{U}$-basis. As in the
quantum case, let us assign different values to the different basis vectors so
the DSD coming from the $U$-basis is: $\left\{  \left\{  \emptyset,\left\{
u_{1}\right\}  \right\}  ,...,\left\{  \emptyset,\left\{  u_{2m}\right\}
\right\}  \right\}  $ and from the $\hat{U}$-basis, the DSD is $\left\{
\left\{  \emptyset,\left\{  \hat{u}_{1}\right\}  \right\}  ,...,\left\{
\emptyset,\left\{  \hat{u}_{2m}\right\}  \right\}  \right\}  $. When the DSD
from the $\hat{U}$-basis is expressed in the computational $U$-basis, then it is
clear that all the intersections of the subspaces from the two DSDs are the
zero space $\left\{  \emptyset\right\}  $ so those DSDs are conjugate.

\section{The basic idea in state reduction (``measurement'')}

\subsection{State reduction = Superposition$^{-1}$}

Superposition adds together definite (eigen) states to give an indefinite
state. State reduction results from an interaction that makes distinctions
between the indefinite states of a superposition. Thus state reduction
``undoes'' what superposition
``does.'' The state reduction takes place
wherever such an interaction occurs which is almost certainly at the quantum
level. Then it must be detected and amplified to the human level to form a
`measurement' in the anthropomorphic sense. 

In the iceberg/partition-lattice representations using support sets, state
reduction is indicated by upward arrows taking an indefinite state to a more
definite state by making distinctions. Support sets get smaller or remain the same. When superposition is misinterpreted as
being like classical wave superposition (definite + definite = definite), then
it is indeed unclear what state reduction is. But when superposition is
interpreted as making an indefinite state out of definite states (definite +
definite = indefinite), then it is easy to see what state reduction does; it
makes distinctions by partly or wholly distinguishing the states in the
superposition state. 

This closes an explanatory circle. Isn't physics supposed to be about masses, forces, and fields? How could the logical notion of distinctions, distinguishings, and differences be a key concept in physics? By interpreting superposition non-classically in terms of indefiniteness, state reduction is understood as the opposite or inverse process of making distinctions. And distinguishing or not distinguishing between the
superposed alternative states is \textit{precisely} the content of the Feynman rules.

\begin{quotation}
	If you could, in principle, distinguish the alternative final states (even
	though you do not bother to do so), the total, final probability is obtained
	by calculating the probability for each state (not the amplitude) and then
	adding them together. If you cannot distinguish the final states even in
	principle, then the probability amplitudes must be summed before taking the
	absolute square to find the actual probability. \cite[Sec. 3.16]
	{feynman:v3mill-ed}
\end{quotation}

\noindent The distinctions or distinguishings between the alternatives has
nothing to do with human observations.

\begin{quotation}
	In other words, the superposition of amplitudes ... is only valid if there
	is no way to know, even in principle, which path the particle took. It is
	important to realize that this does not imply that an observer actually takes
	note of what happens. It is sufficient to destroy the interference pattern, if the path information is accessible in principle from the experiment or even if it is dispersed in the environment and beyond any technical possibility to be recovered, but in principle still ``out
	there.''The absence of any such information is the essential criterion for quantum interference to appear. \cite[484]{zeilinger:1999}
\end{quotation}

\noindent The ``absence of any such
information'' means the absence of distinctions or
distinguishings as in the notion of information-as-distinctions
\cite{ell:NF4IT}.

In his textbooks, Feynman \cite[Sec. 3.3]{feynman:v3mill-ed} always gave
examples at the quantum level of the two cases: distinguishing or not between
the superposed alternative paths. Consider a neutron that is scattering off
the nuclei of atoms in a crystal. If there is no distinguishing which nuclei
was scattered off of, e.g., the nuclei have no spin, then the amplitude for
the neutron to be scattered to some point would be the addition of the
scattering amplitudes off the various nuclei. Since there is no distinguishing
physical event to distinguish between scattering off one nucleus or another,
there is no state reduction in the superposition of the states of scattering
off different nuclei so the amplitudes add. 

But if all the nuclei had spin in,
say, the down direction while the neutron had spin up, then in the scattering
interaction, one of the nuclei might flip its spin which would be the quantum
level physical event to distinguish that trajectory. Then the probability of
the neutron arriving at the given point with its spin reversed (indicating
that a spin flip had occurred) would be the sum of the probabilities (not the
amplitudes) for those distinguished trajectories over all the nuclei. In that
case, the superposition was reduced (the indefinite became definite) and the
nucleus with its spin flipped plays the role of a detector registering a hit.
The spin-state of the nuclei served as a quantum-level measuring apparatus to
distinguish (``measure'') which scattering
trajectory was taken by the neutron to reach the detector. No macroscopic
apparatus was involved in the state reduction (unlike in the `decoherence'
analysis \cite{zurek:decoh}).

\subsection{Illustrating state reduction with Weyl's ``pasta
	machine'' and Feynman's rules}

Hermann Weyl approvingly quoted \cite[255]{weyl:phil} Arthur Eddington who
said that a relativity theorist carries a measuring rod while a quantum
theorist carries a sieve--which Weyl called a ``grating.'' Weyl started with a numerical attribute, e.g.,
$f:U\rightarrow\mathbb{R}$, which defined an inverse-image partition or ``
grating'' or ``aggregate [which] is used in
the sense of `set of elements with equivalence relation.''
\cite[239]{weyl:phil}. Then he, in effect, used the yoga of linearization
so an ``aggregate of $n$ states has to be replaced by an
$n$-dimensional Euclidean vector space'' \cite[256]{weyl:phil} in QM. The notion of a vector space partition or
``grating'' in QM is a ``splitting of the total vector space into mutually orthogonal
subspaces'' so that ``each vector
$\overset{\rightarrow}{x}$ splits into $r$ component vectors lying in the
several subspaces'' \cite[256]{weyl:phil}, i.e., a
direct-sum decomposition of the space. After referring to a partition and its
vector space counterpart, a DSD, as a sieve or grating, Weyl says that
``Measurement means application of a sieve or
grating'' \cite[259]{weyl:phil}, i.e., the making of
distinctions according to which hole in the grating the particle went through.

\begin{figure}[h]
	\centering
	\includegraphics[width=0.9\linewidth]{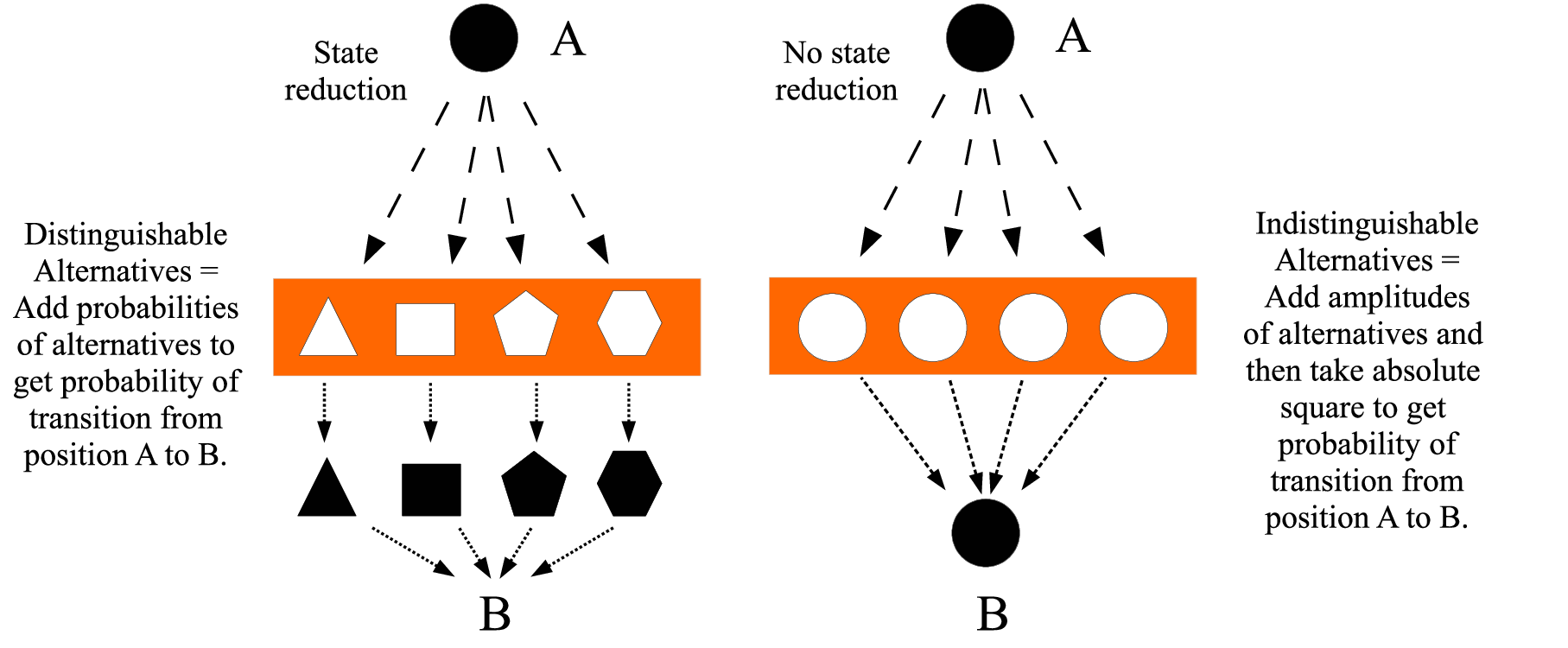}
	\caption{Two cases in Feynman's rule illustrated with Weyl's pasta-machine
		grating imagery}
	\label{fig:two-gratings}
\end{figure}

Weyl's imagery can be illustrated with a ``pasta
machine'' where a ball of pasta (a quantum particle in a
superposition state) has the interaction of going through different holes with
various shapes. The pasta ball can be thought as the indefinite superposition
of the distinct pasta shapes. The two cases of Feynman's rule are illustrated
in Figure \ref{fig:two-gratings}. The left side is the case of the interaction with the pasta grating distinguishing between the different shapes superposed in the pasta ball.

\noindent On the right side is the null grating that makes no distinctions
between the alternative paths from $A$ to $B$ so the amplitudes add and the
absolute square gives the probability of the pasta-ball-particle going from
$A$ to $B$. The pasta-machine imagery gives the basic idea in Feynman's rule
where distinguishable alternative paths implies state reductions and
indistinguishable paths describes unitary evolution by adding the path
amplitudes to get from $A$ to $B$ (in Feynman's path integral formulation of
QM \cite{feynman:path-int}).

The distinguishings have nothing to do with humans. The distinguishings are
the making of distinctions. It has long been suspected that ``
information'' has a fundamental role in QM. This analysis of
state reduction and the definition of information as the quantitative measure
of distinctions, called ``logical entropy''
\ (\cite{manfredi:sp-issue}, \cite{ell:NF4IT}, \cite{ell:qic}), verifies that idea and illustrates it
with upward arrows in the iceberg/partition-lattice diagrams.

\section{The basic idea behind fermions and bosons}

Leibniz and Kant both spelled out a basic idea of classical metaphysics. For
Leibniz, it was the Principle of Identity of Indistinguishables (PII) \cite[22]{Ariew-leib-letters}. If objects are fully definite, then two distinct objects
must have some attribute that one has but not the other. Given two allegedly
different objects, by going down far enough, there must be a distinguishing
attribute, otherwise they are identical. Kant expressed the same idea as the
Principle of Complete Determination.

\begin{quotation}
	\noindent Every thing, however, as to its possibility, further stands under
	the principle of thoroughgoing determination; according to which, among all
	possible predicates of things, insofar as they are compared with their
	opposites, one must apply to it. \cite[B600]{kant:cpr}
\end{quotation}

\noindent In modern terms, classical reality was ``definite
all the way down'' (to paraphrase the joke about
``turtles all the way down\textquotedblright). In the
partition lattice, the discrete partition $\mathbf{1}_{U}$ represents the
classical world so it satisfies the \textit{partition version of Leibniz's
	PII}:

\begin{center}
	If $u$ and $u^{\prime}$ are indistinguishable by $\mathbf{1}_{U}$, i.e.,
	$\left(  u,u^{\prime}\right)  \in\operatorname*{indit}\left(  \mathbf{1}%
	_{U}\right)  $, then $u=u^{\prime}$.
\end{center}

\noindent That is only true for the classical-level partition $\mathbf{1}_{U}%
$; all other partitions contain at least one block with two or more elements
which is the support set of a superposition state and thus non-classical.
However, Figure \ref{fig:twocreationstories} emphasizes that in the Boolean lattice of subsets, there is definiteness all the way down; the elements $a,b,$ and $c$ are always fully
definite. In contrast, for the partition lattice, full classical definiteness
exists only at the top level in the discrete partition $\mathbf{1}_{U}$.

Quantum reality is different from classical reality; it is not definite all
the way down.

\begin{quotation}
	In quantum mechanics, however, identical particles are truly
	indistinguishable. This is because we cannot specify more than a complete set
	of commuting observables for each of the particles; in particular, we cannot
	label the particle by coloring it blue. \cite[446]{sakuri-nap}
\end{quotation}

\noindent Since quantum reality is not definite all the way down, this creates
the possibility of two types of particles:

\begin{enumerate}
	\item Fermions: the type where the existing level of definiteness was
	sufficient to uniquely determine the particle, and
	
	\item Bosons: the type of particle where that limited level of definiteness is
	insufficient to uniquely determine the particle so there could be many
	particles of that type with the same complete description.
\end{enumerate}

\noindent That is the basic idea behind the two types of particles. That basic
idea can be modeled with symmetric and anti-symmetric wave functions, but we
are concerned with the basic idea. There is a sophisticated theorem, the
spin-statistics theorem in quantum field theory \cite{streater-wightman:spin},
that relates the two types of particles to spin, but our purpose is again to
give the \textit{basic idea} behind having two types of particles in a reality
that is not definite all the way down.

Leibniz's PII says that a complete description uniquely determines an entity.
The Pauli Exclusion Principle says that a complete CSCO description uniquely
determines a fermion. Weyl emphasizes that the Pauli Principle is just the
application of the Leibniz PII in a reality that is not definite all the way down.

\begin{quotation}
	\noindent The upshot of it all is that the electrons satisfy Leibniz's
	\textit{principium identitatis indiscernibilium}, or that the electronic gas
	is a ``monomial aggregate'' (Fermi-Dirac
	statistics). ... As to the Leibniz-Pauli exclusion principle, it is found to
	hold for electrons but not for photons. \cite[247]{weyl:phil}
\end{quotation}

For a metaphor, consider postal package addresses that were only definite down
to the street number, i.e., country, state, city, postal code, and street number.
In a neighborhood zoned for single family dwellings, i.e., a ``
fermionic'' neighborhood, the street-number would have a
single family or it would be a vacant lot. In a neighborhood zoned for
multifamily dwellings such as apartment houses, i.e., a ``
bosonic'' neighborhood, the street-number address would be
insufficient to determine the recipient. There could be many recipients
fitting that same street-number address. That difference is the simple result
of the limited addresses. Within the mathematical machinery of QM, the
difference is between anti-symmetric and symmetric wave functions and between
half-integer spin and integer spin, but our goal was to give the basic idea
behind that difference, i.e., in quantum state descriptions not being
``definite all the way down.''

\begin{figure}[h]
	\centering
	\includegraphics[width=0.7\linewidth]{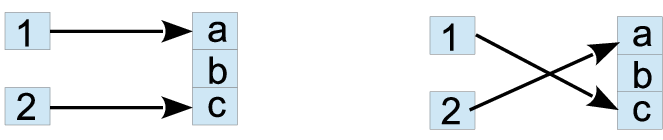}
	\caption{Setup for argument that fermion systems have antisymmtric state vectors}
	\label{fig:fermion-antisymmetry}
\end{figure}

The schematic argument that a complete state description uniquely determines a particle leads to an antisymmetric state vector for a system of indistinguishable fermions is summarized as follows as illustrated in Figure \ref{fig:fermion-antisymmetry}
There are two numerically distinct
but indistinguishable fermions, e.g., electrons, $1$ and $2$, and there are
two ways they could go to states $a$ or $c$. The amplitude $1\rightarrow a
$ and $2\rightarrow c$ is $\left\langle a|1\right\rangle \left\langle
c|2\right\rangle $. If the particles were permuted, then the amplitude is $%
Perm_{1,2}\left\langle a|1\right\rangle \left\langle c|2\right\rangle $.
Since those two arrangements are indistinguishable, the amplitude for that
end state is the sum of the two amplitudes: $\left\langle a|1\right\rangle
\left\langle c|2\right\rangle +Perm_{1,2}\left\langle a|1\right\rangle
\left\langle c|2\right\rangle $. But if we then try to put the two particles
in the same end state $b$, then by assumption that is impossible, i.e., has
amplitude $0$, so $Perm_{1,2}\left\langle b|1\right\rangle \left\langle
b|2\right\rangle =-\left\langle b|1\right\rangle \left\langle
b|2\right\rangle $, and thus in general $Perm_{1,2}\left\langle
a|1\right\rangle \left\langle c|2\right\rangle =-\left\langle
a|1\right\rangle \left\langle c|2\right\rangle $. Those are the basic ideas behind fermions, as opposed to bosons, and the Pauli exclusion principle.

\section{Concluding remarks}

The aim of this paper has been to try to explain in intuitive
(\textit{anschaulich}) terms the basic ideas behind the more puzzling aspects
of quantum mechanics. In summary, here are some of those basic ideas.

\begin{enumerate}
	\item The \textit{basic idea} of superposition:
	
	\begin{enumerate}
		\item as being the flip-side of abstraction--the combination of entities with some similarities and some differences by abstracting away from the differences (making them indefinite) and being definite only on the
		similarities, 
		\item as being the quantum notion (definite + definite = indefinite) unlike the classical notion of superposition (definite + definite = definite), and
		
		\item as being the key feature of quantum states responsible for them forming a vector space.
	\end{enumerate}
	
	\item The \textit{basic idea} behind quantum amplitudes and the Born rule is
	shown by the simple extension of finite probability theory by adding
	superposition events $\Sigma S$ in addition to the usual discrete events $S$. The $n\times n$ matrix representation of superposition events (unlike classical discrete events) has a vector ``square root'' of quantum amplitudes whose square gives the Born rule probabilities for the outcomes of the superposition event.
 	
	\item The \textit{basic idea} of constructing a pedagogical or toy model of
	QM, QM/Sets, by simplifying quantum pure state vectors down to their support
	sets in a vector space over $\mathbb{Z}_{2}$.
	
	\item The \textit{basic idea} of constructing a partition lattice based on a
	pure state whose bottom or indiscrete partition has the only block as the
	support set of the pure state, whose top or discrete partition is the support
	of the corresponding completely decomposed mixed state, and where the
	intermediate partitions consists of the support sets of the mixed states
	resulting from projective measurements of the pure state.
	
	\item The \textit{basic idea} of a projective measurement resulting in a mixed
	state given by the L\"{u}ders mixture operation is shown to be the join
	operation in the partition lattice of support sets.
	
	\item The \textit{basic idea} of a vN Type I process as making distinctions
	and a vN Type II processes as not making distinctions so the measure of the
	indistinctness and distinctness of two quantum states, i.e., the inner
	product, is preserved.
	
	\item The \textit{basic idea} of the partition lattice of a pure state allows
	intuitive pictures of:
	
	\begin{enumerate}
		\item The iceberg/partition-lattice picture of the division between the fully definite classical
		states in the discrete partition at the top and the indefinite states of the
		quantum world represented by the rest of the lattice below the top;
		
		\item Upward moves in the lattice correspond to projective measurements
		(L\"{u}ders mixture operations) of making distinctions;
		
		\item Non-upward (horizontal or downward) non-singular moves in the lattice correspond to
		unitary evolution (not making distinctions); and
		
		\item The classical world emerges from the quantum world by the making of
		distinctions. 
	\end{enumerate}
	
	\item The \textit{basic ideas} of the two-slit experiment as shown by its
	formulation in QM/Sets, e.g., how to answer the question: ``With no detection at the slits, how does the particle get from the two-slit
	screen to the detection wall on the other side of the screen without going through slit 1 or slit 2?''. The table-top ripple-tank demonstration of Case 2 (no detection at the slits) or evocation of the wave-particle complementary magic are attempts to use the (misleading) \textit{classical} (definite + definite = definite) notion of superposition of waves instead of the (below-the-water) same-math evolution of the quantum (definite + definite = indefinite) superposition of states.
 	
	\item The \textit{basic idea} of linearization to establish a dictionary
	relating concepts of sets math, e.g., real-valued numerical attributes $f$ or
	partitions $f^{-1}$, and the corresponding Hilbert space QM-analogs, e.g.,
	Hermitian operators $\widehat{f}$ or direct sum decompositions $\widehat{f^{-1}}$.
	
	\item The \textit{basic idea} of non-commutativity of operators in QM was
	analyzed showing that this was more an aspect of matrix math in vector spaces
	so it is not a unique characteristic of QM and this was illustrated by giving
	a case of conjugacy arising from Fourier-like transforms in QM/Sets for $\mathbb{Z}_{2}^{2m}$ with $m>1$.
	
	\item The \textit{basic idea} of state reduction (``
	measurement\textquotedblright) as being the inverse of superposition where
	superposition arises by making indefinite the differences between eigenstates
	and state reduction results from an interaction that distinguishes between the
	superposed states (or paths)--where the two vN cases of a distinguishing and a
	non-distinguishing interaction were intuitively illustrated by Weyl's
	``pasta machine'' and the Feynman rules. 
	
	\item The \textit{basic idea} at the logical level is distinctions versus
	indistinctions, differences versus similarities, distinguishings versus
	indistinguishings--all represented at the logical level in the logic of
	partitions \cite{ell:lop-book}.
	
	\item The \textit{basic idea} of information-as-distinctions is the
	quantitative version of the logic of partitions \cite{ell:NF4IT}. Since
	indistinctions (i.e., superpositions) and distinctions (state reduction =
	superposition$^{-1}$) have an ontic role in QM, this verifies the old idea
	that the quantitative version of distinctions, i.e., information, has an ontic
	role in QM.
	
	\item The \textit{basic idea} of the two types of particles, fermions and
	bosons, arises from the contrast between classical reality as being definite
	all the way down and quantum reality as being definite only down to a certain
	level, i.e., as given by a complete set of commuting observables (CSCO), so
	some particles will be uniquely identified by that limited degree of
	definiteness (fermions) and other numerically distinct particles can all have
	the same limited state description (bosons).
\end{enumerate}

In the full Hilbert space machinery of QM, all these `gears' mesh together
beautifully to make our most highly verified physical theory. In that sense,
the theory is not the problem; the problem is how to intuitively conceptualize
the underlying physical reality. Our approach to understanding that underlying
reality has been to break down the machinery into basic ideas that can be
understood in an intuitive manner.


\section*{Statements and Declarations}

\begin{itemize}
\item Funding: None.
\item Conflict of interest/Competing interests: None.
\end{itemize}

\bigskip


\end{document}